%% file: Spin-S_Kitaev_Ladder.tex
\documentclass[aps,prb,reprint,superscriptaddress,amsmath]{revtex4-2}
\usepackage{graphicx}
\usepackage{enumerate}
\usepackage{refcount}
\usepackage{fmtcount}
\usepackage{amssymb}
\usepackage{dsfont}
\usepackage{hyperref}
\usepackage[usenames,dvipsnames]{color}
\usepackage[normalem]{ulem}
\usepackage{enumitem}

\usepackage{float}

\usepackage{array}
\usepackage{multirow}
\usepackage{physics}

\hypersetup{pdfstartview=FitH,pdfpagemode=UseOutlines,colorlinks,
citecolor=blue,linkcolor=blue,urlcolor=blue}

\graphicspath{{.}{./Folder_with_figures/}} 

\newcommand{\ox}{e^{i\pi S^x}}
\newcommand{\oy}{e^{i\pi S^y}}
\newcommand{\oz}{e^{i\pi S^z}}

\definecolor{aaron}{rgb}{0.6, 0.6, 0.8}

\begin{document}

\title{\texorpdfstring{Phase diagrams of spin-$S$ Kitaev ladders}{Phase diagrams of spin-S Kitaev ladders}}

\author{Yushao Chen}
\email[]{ychen2@perimeterinstitute.ca}
	\affiliation{Perimeter Institute for Theoretical Physics, Waterloo, Ontario N2L 2Y5, Canada}
 	\affiliation{Department of Physics and Astronomy, University of Waterloo, Waterloo, Ontario, Canada N2L 3G1}
 \author{Yin-Chen He}
	\affiliation{Perimeter Institute for Theoretical Physics, Waterloo, Ontario N2L 2Y5, Canada}
\author{Aaron Szasz}
	\affiliation{Perimeter Institute for Theoretical Physics, Waterloo, Ontario N2L 2Y5, Canada}
	\affiliation{Computational Research Division, Lawrence Berkeley National Laboratory, Berkeley, California 94720, USA}

\date{\today}

\begin{abstract}
We investigate the ground states of spin-$S$ Kitaev ladders using exact analytical solutions (for $S=1/2$), perturbation theory and the density matrix renormalization group (DMRG) method.
We find an even-odd effect: in the case of half-integer $S$, we find phases with spontaneous symmetry breaking (SSB) and symmetry-protected topological (SPT) order; for integer $S$, we find SSB and trivial paramagnetic phases. We also study the transitions between the various phases; notably, for half-integer $S$ we find a transition between two distinct SPT orders, and for integer $S$ we find unnecessary first order phase transitions within a trivial phase.
\end{abstract}

\maketitle


\section{Introduction}

Quantum spin liquids (QSL) are highly entangled phases of matter with no magnetic order~\cite{Balents2010,zhou2017quantum,savary2016quantum}.
Despite decades of study, many mysteries remain, including theoretical questions about the classification of phases and their properties, as well as whether and where QSLs will arise in simple microscopic models or indeed in real materials.
Kitaev spin liquids (KSL) \cite{KITAEV20062,TREBST20221,Hermanns2018,Takagi2019_KSL_review} provide a particularly promising avenue of study, due to the exact solvability of the spin-$1/2$ Kitaev honeycomb model and the existence of possible  experimental realizations of Kitaev interactions due to spin-orbit coupling~\cite{Jackeli2009,WitczakKrempa2014,Winter2016,Jang2019,Motome_2020_JPCM} in materials such as $\alpha$-Li$_2$IrO$_3$~\cite{Singh2012,Williams2016,Tsirlin2022} and $\alpha$-$\mathrm{RuCl}_{3}$~\cite{Plumb2014,Kim2015,banerjee2016proximate,Kim2016,Cookmeyer2018,Kasahara2018,Yokoi2021,Loidl_2021}.

Recently, studies have also focused on higher-spin analogs of KSLs. 
Spin-1 and spin-3/2 Kitaev honeycomb models have been found~\cite{Khait2021spin1, koga2018_S1, Dong2020_spin1, HuiKe2022_S32, zhu2020magnetic, consoli2020heisenberg, manmana2013topological} to provide microscopic descriptions of materials such as $\mathrm{NiI}_2$, $\mathrm{Li}_3 \mathrm{Ni}_2 \mathrm{Bi} \mathrm{O}_6$~\cite{stavropoulos2019microscopic}, $\mathrm{CrGeTe}_3$, $\mathrm{CrI}_3$~\cite{Stavropoulos2021_CrI,  Xu2018_CrI} and $\mathrm{CrSiTe}_3$~\cite{xu2020possible}.
However, in contrast to spin-$1/2$, the higher-spin Kitaev honeycomb models are not exactly solvable, and it is necessary to turn to numerical simulation.
Furthermore, such simulations become exponentially more computationally expensive as the spin, and thus the dimension of the local Hilbert space, increases. 
Consequently, despite numerical studies using methods such as exact diagonalization~\cite{bradley2022instabilities}, parton mean-field theory~\cite{HuiKe2022_S32} and tensor networks~\cite{Lee2020, lee2020magnetic}, many questions remain.

\begin{figure}[t]
    \centering
    \includegraphics[width=0.9\linewidth]{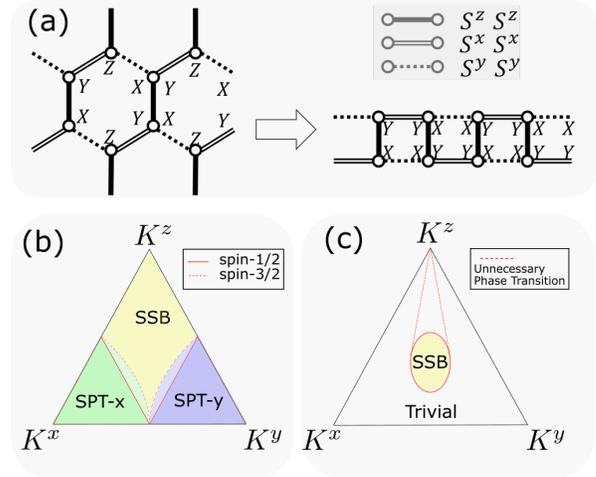}
    \caption{(a) Definition of Kitaev model on a honeycomb lattice and a two-leg ladder: solid, double, and dashed lines indicate spin interactions as shown in the key, with respective strengths $K_z$, $K_x$, and $K_y$. There is a local symmetry for each plaquette given by a product of local operators $X=e^{i\pi S^x}$, etc. along the interior of each plaquette. (b) Phase diagram of half-integer spin Kitaev ladders; red lines indicate the critical points. (c) Phase diagram of integer spin Kitaev ladders. Dashed lines indicate the first-order phase transition lines of spin-$1$ ladders. For higher spins, the unnecessary phase transition locations may change.}
    \label{fig:demo_and_pd}
\end{figure}

To make progress on this difficult problem, we consider the Kitaev model on a two-leg ladder, with $S=1/2$, $1$, $3/2$, and $2$.
By understanding the correspondence between our results for the spin-1/2 ladder and the known behavior of the spin-1/2 honeycomb model, we may use the results for higher-spin models on the ladder to gain insight by analogy into the higher-spin honeycomb models, while solving a more numerically approachable problem.
The major benefit, numerically, of studying two-leg ladders rather than the full honeycomb model is the applicability of the density matrix renormalization group (DMRG) technique~\cite{White1992,White1993};
DMRG is a variational algorithm optimizing one-dimensional (1D) quantum states within the class of matrix product states (MPSs)~\cite{Schollwock2011}.
MPSs can provide approximations of mildly entangled states in 1D with high accuracy due to efficient truncation over the high-dimensional Hilbert space, while guaranteeing a low computational cost.
It is common to use DMRG to simulate 1D or quasi-1D quantum systems on either finite, open chains (fDMRG) or infinite, translationally invariant chains (iDMRG); we use both approaches in this work.
While DMRG is most effective for studying gapped ground states, for which the area law guarantees low entanglement, even for gapless regimes, where the ground states have long-range correlations, DMRG still provides precise results via scaling with the MPS bond dimension, a tunable parameter.

We summarize our main results in Fig.~\ref{fig:demo_and_pd}. For half-integer spin-$S$ Kitaev ladders, the phase diagram is composed of three regimes, a spontaneous symmetry breaking (SSB) phase  and two symmetry-protected topological (SPT) phases. For integer spin-$S$ Kitaev ladders, there is an SSB phase around the isotropic point, surrounded by a trivial phase.

The layout of the remainder of the paper is as follows.
In Sec \ref{sec:Sec2} we define the Kitaev model and discuss the generalization of symmetries of the spin-$1/2$ model to arbitrary spin-$S$ Kitaev ladders. We also present the results of perturbation theory in two distinct anisotropic limits and for integer and half-integer spins.
In Sec \ref{sec:results}, we show detailed results of both analytical calculation and DMRG simulations for spin-$1/2$, spin-$1$, spin-$3/2$ and spin-$2$ Kitaev ladders; for each case we demonstrate consistency between perturbation theory and numerical results.
In Sec \ref{sec:discuss}, we summarize our findings and discuss the significance of the results.


\section{The model\label{sec:Sec2}}
We study the Kitaev model on a two-leg ladder.  For any spin size, $S$, the Hamiltonian is
\begin{equation}
    H(\mathbf{K}) = \sum_{\langle ij \rangle} K_\gamma {S}^\gamma_{i} {S}^\gamma_{j},
\end{equation}
where the parameter $\mathbf{K}$ has three components, $\mathbf{K} = (K_x, K_y, K_z)$, and each nearest-neighbor bond $\langle ij \rangle$ has a single spin component $\gamma = x,y,$ or $z$ as shown in Fig.~\ref{fig:demo_and_pd}(a). 
The ladder model on the right of Fig.~\ref{fig:demo_and_pd}(a) is derived from the standard Kitaev honeycomb model simply by applying periodic boundary conditions (in the vertical direction) on the width-2 strip of the honeycomb model on the left of Fig.~\ref{fig:demo_and_pd}(a).

\subsection{Global and local symmetries\label{subsection: symmetries}}
For higher spin, the spin-$1/2$ anticommutation relations, $\{S^\alpha,S^\beta\}=\delta^{\alpha \beta}$, do not directly generalize.  
Instead, we have $\{S^\alpha,e^{i \pi S^\beta}\}=0, \forall \alpha \neq \beta$, and $\left[ S^\alpha,e^{i \pi S^\alpha}\right]=0$~\cite{Baskaran2008_spinS_KSL, spin1in1DDiptimanSen, ThermoIntKMDSen}; in other words, conjugation by, for example, $e^{i\pi S^x}$ acts as a $\pi$ rotation around $x$, thus transforming $S^x\mapsto S^x$, $S^y\mapsto -S^y$ and $S^z\mapsto-S^z$.
It follows immediately that
\begin{equation}
    [e^{i\pi S^\alpha}\otimes e^{i\pi S^\alpha}, S^\beta\otimes S^\beta] = 0\label{eq:spin_rot_com}
\end{equation}
$\forall \alpha$, $\beta$ and for any size spin. 
Using this relation, we can determine the global and local symmetries of our Hamiltonian.

We first consider the global symmetries.  
Every term in the model is of the form $S_i^\gamma S_j^\gamma$ for $\gamma\in\{x,y,z\}$, and by Eq.~\eqref{eq:spin_rot_com}, each such term commutes with $e^{i\pi S_i^{\gamma'}}e^{i\pi S_j^{\gamma'}}$ for all $\gamma'$.
As a result, the Hamiltonian is invariant under conjugation by the global on-site symmetry operators $\Sigma^\gamma = \prod_n e^{i\pi S^{\gamma}_{n}}, \forall \gamma \in \{x, y, z\}$.
Furthermore, for the $2$-leg Kitaev ladder in particular, the product of $e^{i \pi S^z}$ along each individual leg of the ladder is an extra global symmetry, which we denote by $\Sigma^Z_u$ for the upper leg and $\Sigma^Z_l$ for the lower leg.

There are also local symmetries.  On a lattice with coordination number 3, any self-avoiding closed loop $\Gamma$ will cover two of the three bonds connected to each site on the loop; letting $\gamma_i$ be the spin-component ($x$, $y$, or $z$) for the bond on site $i$ that is not covered by the loop, the operator $W_\Gamma=\prod_i e^{i \pi S^{\gamma_i}_i}$ commutes with $H$; this follows from a generalization of Eq.~\eqref{eq:spin_rot_com}, that $[e^{i\pi S_i}\otimes e^{i\pi S_j}, S_k\otimes S_k] = 0$ if $i\neq k$ and $j\neq k$.
Each such $W_\Gamma$ is thus a local symmetry of the Hamiltonian, but many are redundant with each other.  A non-redundant set of local symmetries is given by taking all paths $\Gamma$ that are individual lattice plaquettes.

On the honeycomb lattice, this construction gives the usual local symmetry operators on each hexagonal plaquette, as illustrated in Fig.~\ref{fig:demo_and_pd}(a).  Each $W$ has eigenvalues $\pm 1$, and a Hamiltonian with $2N$ spins will have $N$ hexagonal plaquettes, so the Hamiltonian can be block-diagonalized into $2^N$ eigenspaces given by fixing the value of each $W$; each block has dimension $(2S+1)^{2N}/2^N$.

On the ladder, the same construction gives an independent local symmetry on each square plaquette, each of the form $D=e^{i\pi S^x}e^{i\pi S^x}e^{i\pi S^y}e^{i\pi S^y}$, where the $x$ rotations act on the sites linked by a $y$ bond and vice-versa, as shown in Fig.~\ref{fig:demo_and_pd}(a).  There are still two spins per plaquette, so fixing the value of each $D$ again divides the Hamiltonian into $2^N$ blocks of dimension $(2S+1)^{2N}/2^N$.  This analogous behavior of local symmetries on the ladder and the honeycomb is a strong reason to believe that our present study of higher-spin ladder models will also provide insight into the higher-spin honeycomb.

We summarize the symmetries of spin-$S$ ladders as follows:

\paragraph*{Global symmetries}
 The global symmetry group is generated by $\Sigma^X, \Sigma^Y$ and $\Sigma^Z_u$. 
 They generate a $\mathbb{Z}_2 \cross \mathbb{Z}_2 \cross \mathbb{Z}_2$ group on ladders of $4N$ spins.
To be explicit, the full symmetry group is:
\begin{equation}
    G = \left\{ \mathbf{Id},  \Sigma^X, \Sigma^Y, \Sigma^Z, \Sigma^Z_u, \Sigma^Z_l, \Sigma^X_u\Sigma^Y_l, \Sigma^X_l\Sigma^Y_u\right\}\label{eq:sym_group_full}
\end{equation}
where $\Sigma^X_u\Sigma^Y_l$ is defined by acting with $e^{i\pi S^x}$ on all sites of the upper leg and $e^{i\pi S^y}$ on the lower leg, and likewise for $\Sigma^X_l\Sigma^Y_u$.

\paragraph*{Local symmetries} 
The loop operators $D_n$ defined on the smallest square plaquettes are the local symmetries. For a periodic ladder of $4N$ spins, there will be $2N$ independent local symmetries $D_n$.
Each has eigenvalues $\pm1$, so when we fix the value of each one, the dimension of Hilbert space will be reduced by $2^{2N}$.

\subsection{Perturbation theory in anisotropic limits}
The spin-1/2 Kitaev honeycomb model is exactly solvable, and the same is true of the spin-1/2 ladder \cite{Minakawa2019PT}.  
However, the higher-spin models, both on the honeycomb and on the ladder, are no longer exactly solvable.  
Furthermore, for higher-spin-$S$ the local Hilbert space dimension is larger, while the local symmetries discussed above provide only the same reduction as in the spin-1/2 case. 
However, we can use perturbation theory to  find simpler effective models in the  anisotropic limits $K_x \gg K_y, K_z>0$ and $K_z \gg K_x, K_y>0$. (The case of large $K_y$ is equivalent to large $K_x$ by symmetry.)

To find an effective model using perturbation theory, a Hamiltonian $H$ is split as $H=H_0 + H^\prime$, with $H^\prime$ small. 
The degrees of freedom for the effective model are given by the (often highly degenerate) ground state subspace of $H_0$, $\mathcal{H}_{\textbf{GS}}$, while the interactions are determined by matrix elements of powers of $H^\prime$. 
For Kitaev ladders in the limits where either $K_z$ (``$Z$-limit'') or $K_x$ (``$X$-limit'') dominates, the corresponding term serves as $H_0$ and the other two terms serve as the perturbation:
\begin{align}
    H_{0, \textbf{Z}} &= K_z h^z 
    ,\,\, H^\prime_{\textbf{Z}} = K_x h^x + K_y h^y,\\
    H_{0, \textbf{X}} &= K_x h^x
    , \,\,H^\prime_{\textbf{X}} = K_z h^z + K_y h^y,
\end{align}
where $h^\gamma = \sum_{\expval{i,j}_\gamma} S^\gamma_i S^\gamma_j$. 

\begin{figure}
    \centering
    \includegraphics[width=.9\linewidth]{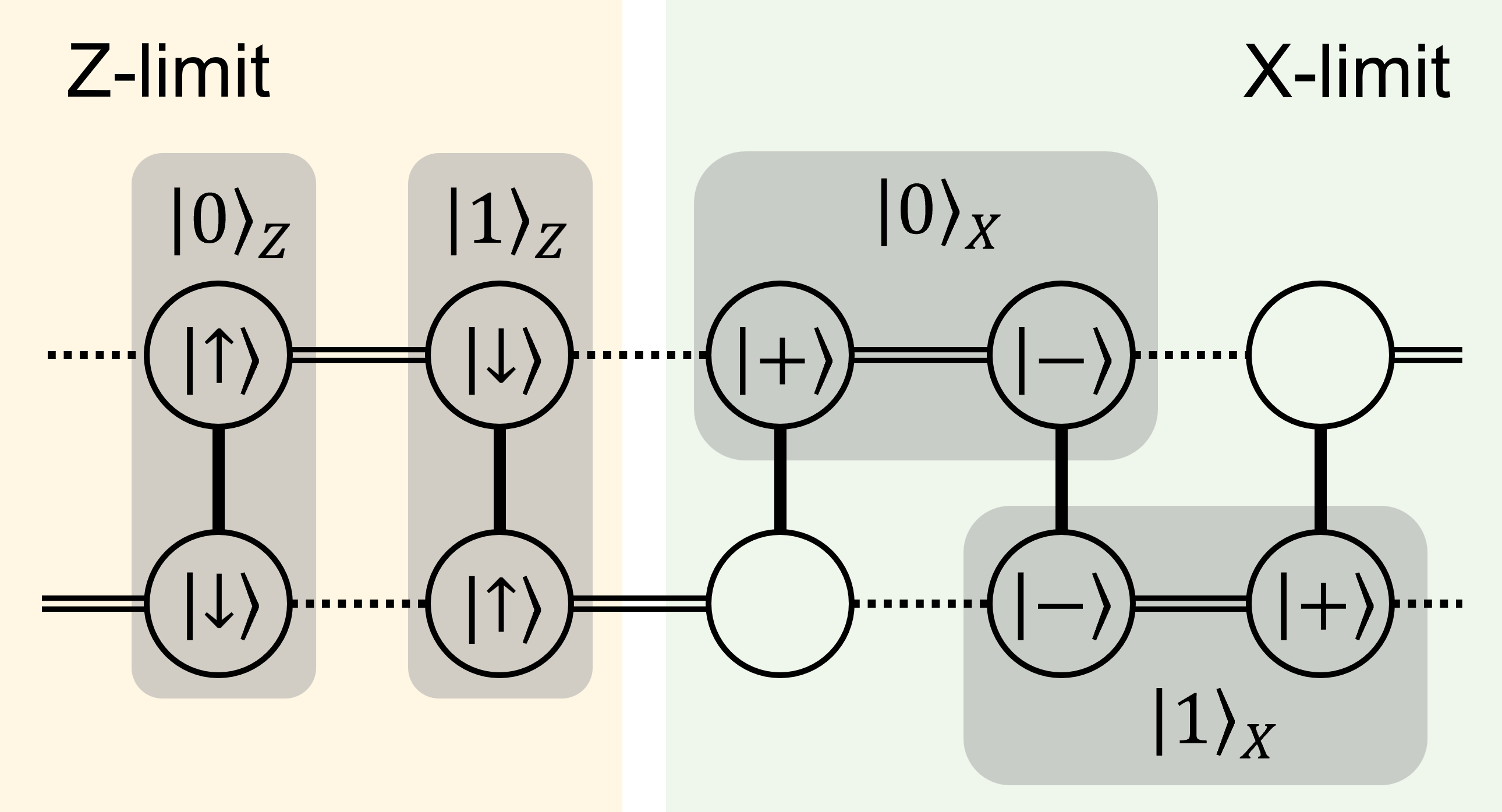}
    \caption{Dimer illustration in both anisotropic limits. Here we define $\ket{\uparrow}$ and $\ket{\downarrow}$ to be the eigenstates of $S^z$ s.t. $S^z\ket{\uparrow} = S\ket{\uparrow}, S^z\ket{\downarrow} = -S\ket{\downarrow}, $ and $\ket{\pm}$ the corresponding eigenstates of $S^x$.}
    \label{fig:xlimit}
\end{figure}

In either limit, the ground subspace of $H_0$ defines an effective spin-1/2 chain.
$H_0$ acts independently on dimers corresponding to the strongest bond, and as depicted in the shaded boxes of Fig.~\ref{fig:xlimit}, there are two degenerate local ground states for each dimer, defining the effective spin-1/2 degree of freedom: $\ket{0}_k = \ket{S^k=+S}\ket{S^k=-S}$, and $\ket{1}_k = \ket{S^k=-S}\ket{S^k=+S} $; $k=x,z$ corresponds to the limit taken and $\ket{S^k=\pm S}$ are the $\pm S$ eigenstates of $S^k$. 
The Hamiltonian for this effective spin-1/2 chain is then determined by the powers of $H^\prime$ that act nontrivially in $\mathcal{H}_{\textbf{GS}}$; we present the procedure for finding the effective model in detail in Appendix~\ref{appendix:PT}.  In practice we find distinct effective models in four cases: in the $X$- and $Z$-limits with half-integer and integer spin.

We denote the Pauli operators for the effective model as $\tau$, with the usual definitions, e.g. $\tau^z \ket{0} = \ket{0}, \tau^z \ket{1} = -\ket{1}$.  Since the Hilbert space dimension for each dimer is reduced from $(2S+1)^2$ to 2, in general there are many operators in the original model corresponding to each Pauli operator in the effective model.  Representations in which the operator in the original model acts as $e^{i\pi S^k}$ on each spin, presented in Table~\ref{table:localeff}, are particularly useful, as they make clear the relationship between the $\tau$ operators and the symmetries of the original model.  Indeed, all global and local symmetries take a simple form in the effective model in each of the four cases, as we summarize in Table~\ref{table:eff}.

\begin{table}[htbp]
\renewcommand{\arraystretch}{1.2}
\begin{tabular}{ c  | >{\centering\arraybackslash}p{1.8cm}| >{\centering\arraybackslash}p{1.8cm} | >{\centering\arraybackslash}p{1.8cm} |>{\centering\arraybackslash}p{2.1cm} }
  \toprule
        \multirow{2}{*}{ }& \multicolumn{2}{c|}{Half-Integer}  & \multicolumn{2}{c}{Integer} \\  \cline{2-5}
        & $X$-lim  & $Z$-lim  & $X$-lim  & $Z$-lim  \\  \hline
   $ \tau^x$ & $\oz \otimes \oz$ & $\ox \otimes \ox$ &$\oz \otimes \oz$ &$-\ox \otimes \oy$ \\  \hline
   $ \tau^y$ & $\oz \otimes \oy$ & $\ox \otimes \oy$ & - & - \\  \hline
   $ \tau^z$ & $\ox \otimes \text{Id}$ \text{ }\text{ }& $\oz \otimes \text{Id}$ \text{ }\text{ }& - & - \\  \hline
  \hline
\end{tabular}
 \caption{Mapping from the original local operators to the effective model.
 Some spaces are left blank since there is no simple way to represent those effective operators in terms of $\ox$, $\oy$, and $\oz$.
 For the $Z$-limit of integer system, $-\ox \otimes \oy$ and $\ox \otimes \ox$ both map to $\tau^x$}
 \label{table:localeff}
\end{table}

We now present the derived effective spin-chain models in the two limits and with integer and half-integer spin, also summarized in Table~\ref{table:eff}.  
We will see that the effective models are essentially given by a summation of the local symmetries.  We comment on this surprising correspondence in Sec~\ref{sec:Heff_D}.

\begin{table}[htbp]
\renewcommand{\arraystretch}{1.2}
\begin{tabular}{ c  | >{\centering\arraybackslash}p{1.5cm}| >{\centering\arraybackslash}p{1.5cm} | >{\centering\arraybackslash}p{1.5cm} |>{\centering\arraybackslash}p{1.5cm} }
  \toprule
        \multirow{2}{*}{ }& \multicolumn{2}{c|}{Half-Integer}  & \multicolumn{2}{c}{Integer} \\  \cline{2-5}
        & $X$-lim  & $Z$-lim  & $X$-lim  & $Z$-lim  \\  \hline
   $h_{\text{eff}}$ & $\tau^z\tau^x\tau^z$ & $\tau^y\tau^y$ &$\tau^x$ &$\tau^x+\tau^x\tau^x$ \\  \hline
   $D_n$ & $\tau^z\tau^x\tau^z$ & $\tau^y\tau^y$ &$\tau^x$ &$\tau^x\tau^x$ \\  \hline
   $\Sigma^X$ & $-\text{Id}$ & $X$ &$\text{Id}$ &$X$ \\  \hline
   $\Sigma^Y$ & $X$ & $X$ &$X$ &$X$ \\  \hline
   $\Sigma^Z$ & $X$ & $-\text{Id}$ &$X$ &$\text{Id}$ \\  \hline
   $\Sigma^Z_l$ & $X_{\text{odd}}$ & $Z$ &$X$ &$\text{Id}$ \\  \hline
   $\Sigma^Z_u$ & $X_{\text{even}}$ & $-Z$ &$X$ &$\text{Id}$ \\  \hline
  \hline
\end{tabular}
 \caption{Effective Hamiltonian and effective symmetries given by perturbation theory. $X$ and $Z$ are the global on-site operators in the effective model, e.g. $X = \otimes_i \tau^x_i$.}
 \label{table:eff}
\end{table}

As we can see from Table~\ref{table:eff}, in the half-integer-spin case ($Z$-limit) the effective operator $\tau^y$ anti-commutes with most of the global symmetries, and thus can serve as an order parameter for SSB. In other words, $\left\langle\ox \otimes \oy\right\rangle$ on a $Z$-bond as shown in the inset of Fig.~\ref{fig:spin1half}(a) is a good order parameter.  On the other hand, the operator $\ox \otimes \oy$ commutes with all global symmetries in the integer-spin case and thus cannot be used as an order parameter.  Instead, we can use $\langle S^x S^y\rangle$, which anti-commutes with global symmetry operators.

\subsubsection{\texorpdfstring{Half-integer spin-$S$}{Half-integer spin-S}\label{sec:half_X_lim}}
\paragraph*{$X$-limit} When $K_x$ dominates, the ground subspace is composed of dimers on the original $X$-bonds as indicated in the right part of  Fig.~\ref{fig:xlimit},  and
\begin{equation}
    H_{\text{eff}}(K) = \alpha_{S,x}(K)\sum_{i = 1}^{2N} \tau^z_{i} \tau^x_{i+1} \tau^z_{i+2}.
\end{equation}
where $\alpha$ is a coefficient depending on $S$ and $K$.

This Hamiltonian is the cluster model~\cite{Son2012}, a prototypical model of symmetry protected topological (SPT) order \cite{spt, sptFrankPollmann, sptXieChen}.  
The symmetry group protecting it is $\mathbb{Z}_2 \cross \mathbb{Z}_2$, with the two copies of $\mathbb{Z}_2$ given by $X_\text{even} = \prod_{i} \tau^x_{2i}$ and $X_\text{odd} = \prod_{i} \tau^x_{2i+1}$.

Recall that the original model has a symmetry group $\mathbb{Z}_2 \cross \mathbb{Z}_2 \cross \mathbb{Z}_2$ generated by $\Sigma^X$, $\Sigma^Y$, and $\Sigma^Z_u$.
The full symmetry group is given in Eq.~\eqref{eq:sym_group_full}. 
Several $\mathbb{Z}_2 \cross \mathbb{Z}_2$ subgroups of the full symmetry group map to $X_\text{even}\cross X_\text{odd}$, the protecting symmetry of the effective cluster model.  One obvious candidate $\mathbb{Z}_2 \cross \mathbb{Z}_2$ subgroup is $\Sigma^Z_u \times \Sigma^Z_l$, as $\Sigma^Z_u$ and $\Sigma^Z_l$ act in the effective model as $X_\text{even}$ and $X_\text{odd}$ respectively.  However, $\Sigma^Y$ acts in the effective model as the global $X$ symmetry, so $\Sigma^X_u \Sigma^Y_l$ also acts as $X_\text{odd}$ and $\Sigma^X_l \Sigma^Y_u$ acts as $X_\text{even}$.  Thus two additional candidate $\mathbb{Z}_2 \cross \mathbb{Z}_2$ subgroups are $\Sigma^X_u \Sigma^Y_l \cross \Sigma^Z_u$ and $\Sigma^X_l \Sigma^Y_u \cross \Sigma^Z_l$.  (These two $\mathbb{Z}_2 \cross \mathbb{Z}_2$ subgroups are equivalent up to a glide symmetry of the Hamiltonian.)

As noted above, the large $K_y$ limit is equivalent to this large $K_x$ limit by symmetry, but it is worth briefly commenting on the difference in the mapping of symmetries from the original model to the effective model.  In this case, the roles of $\Sigma^X$ and $\Sigma^Y$ in the effective model are swapped -- the former acts as the global $X$ symmetry, while the latter acts as the identity.  Additionally, the images of $\Sigma^Z_u$ and $\Sigma^Z_l$ are swapped: $\Sigma^Z_u \cross \Sigma^Z_l \mapsto X_\text{odd} \cross X_\text{even}$. (More precisely, the mapping from $(u,l)$ to $(\text{even},\text{odd})$ vs $(\text{odd},\text{even})$ depends on how we choose our unit cell, but for any consistent choice, the $X$- and $Y$-limits will map in opposite ways.)  Most importantly, the candidate protecting $\mathbb{Z}_2 \cross \mathbb{Z}_2$ symmetry groups for the cluster model in this limit are $\Sigma^Z_u \cross \Sigma^Z_l$, $\Sigma^X_l \Sigma^Y_u \cross \Sigma^Z_u$, and $\Sigma^X_u \Sigma^Y_l \cross \Sigma^Z_l$.  

To summarize, the SPT order in the effective model, and hence in the extreme anisotropic limit of the original Kitaev ladder, is protected in both $X$- and $Y$-limits by $\Sigma^Z_u \cross \Sigma^Z_l$.  In the $X$-limit $\Sigma^X_u \Sigma^Y_l \cross \Sigma^Z_u$ and $\Sigma^X_l \Sigma^Y_u \cross \Sigma^Z_l$ also protect the SPT.  In the $Y$-limit the latter subgroups are replaced by $\Sigma^X_l \Sigma^Y_u \cross \Sigma^Z_u$ and $\Sigma^X_u \Sigma^Y_l \cross \Sigma^Z_l$.  In Section~\ref{sec:results} below, we show that the latter symmetry groups protect SPT order even away from the extreme anisotropic limit; since the protecting symmetry groups are different, we conclude that there are two distinct SPT phases in the $X$- and $Y$-limits.

\paragraph*{$Z$-limit} When $K_z$ dominates, the ground subspace is composed of dimers on the original $Z$-bonds as indicated in the left part of Fig.~\ref{fig:xlimit}, and
\begin{equation}
    H_{\text{eff}}(K) = \alpha_{S,z}(K) \sum_{i = 1}^{2N} \tau^y_{i} \tau^y_{i+1}.
\end{equation}
This Hamiltonian is an Ising model whose ground states have an order parameter $\tau^y$, corresponding to the operator $\ox \otimes \oy$ and breaking the symmetries $\Sigma^X, \Sigma^Y, \Sigma^Z_u$ and $\Sigma^Z_l$ in the original model.

\subsubsection{\texorpdfstring{Integer spin-$S$}{Integer spin-S}}
\paragraph*{$X$-limit} The configuration of the ground subspace does not change with $S$, so it again consists of dimers on the original $X$-bonds.
However, the effective Hamiltonian is different from that of the half-integer case, 
\begin{equation}
    H_{\text{eff}}(K) = \alpha_{S,x}(K)\sum_{i = 1}^{2N} \tau^x_{i},
\end{equation}
which gives rise to a trivial ground state connected to a product state of dimers.

\paragraph*{$Z$-limit} 
In the final case, 
\begin{equation} \label{eq:unnecessaryPT}
    H_{\text{eff}}(K) = \sum_{i = 1}^{2N} \alpha_{S,z}(K) \tau^x_{i}  + \beta_{S,z}(K) \tau^x_{i}\tau^x_{i+1}.
\end{equation}
There are two different terms in this effective Hamiltonian, with independent coefficients, allowing for first-order phase transitions between product states as will be discussed later.

\paragraph*{SSB} 
In both limits for integer $S$, there is no term favoring spotaneous symmetry breaking. 
Note, however, that as reported below we find from numerical simulation that an SSB phase does appear for integer $S$ when away from the anisotropic limits.

\subsubsection{Effective models and local symmetries\label{sec:Heff_D}}
Comparing the first two rows in Table~\ref{table:eff}, we see that in each case the lowest-order effective model is given by a summation of the local symmetries (with an extra commuting term in one case).  This correspondence is very useful from a practical perspective: by solving a simple commuting Hamiltonian in the effective model, we immediately get the ground state flux configuration in the original model even in higher-spin cases where Lieb's theorem~\cite{lieb2004flux} may not apply.~\footnote{As a result, we conclude for example that the flux configuration used in Ref.~\cite{WU20123530} is not correct.}

If the perturbation theory produces a unique ground state by selecting a flux configuration, it must also be the case that the ground subspace of $H_0$, $\mathcal{H}_{\text{GS}}$, contains only one state for any given flux configuration.  In other words, the intersection of $\mathcal{H}_{\text{GS}}$ with the subspace of $\mathcal{H}$ corresponding to each diagonal block of $H$ must be one-dimensional, i.e. consist of a single state.  This is indeed correct in the $X$-limit for both integer and half-integer spins.  In the $Z$-limit, only flux configurations with an even number of $D=+1$ appear in the effective model, so in fact the intersection of $\mathcal{H}_{\text{GS}}$ with each allowed fixed-flux subspace is two-dimensional; in the half-integer case, there are consequently two ground states of the perturbative model, giving rise to the $\mathbb{Z}_2$ SSB order, while in the integer case the unique ground state in the two-dimensional intersection is selected by the extra $\tau^x$ in the effective model.

\subsubsection{Comparison with honeycomb lattice perturbation theory}

Similar perturbation theory analysis has been carried out for the Kitaev honeycomb model, both with spin-1/2~\cite{KITAEV20062}  and higher-spin~\cite{Minakawa2019PT}.  On the honeycomb, there is only one perturbative limit to consider, since the cases of large $K_x$, large $K_y$, and large $K_z$ are related by symmetry.  In any of these limits, the half-integer-spin model reduces to the toric code, while the integer-spin model reduces to a disconnected spins in a magnetic field, giving a trivial ground state.~\cite{Minakawa2019PT}  
Evidently, these results on the spin-$S$ honeycomb model match up well with our $X$-limit results: the cluster model is a one-dimensional analogue of the toric code model, while for integer spin we also find the effective Hamiltonian to be a trivial paramagnet.  On the other hand, the $Z$-limit in our perturbation theory does not appear to correspond with any known behavior on the honeycomb lattice.  
Later, in Sec.~\ref{sec:discuss} below, we further discuss the correspondence between the ladder and honeycomb models.


\section{Results\label{sec:results}}
\subsection{Phase diagram of spin-1/2 Kitaev ladder\label{sec:results_spin_half}} 

For the case of spin-1/2, we employ both analytical and numerical approaches to conclusively determine the phase diagram.
Our results are consistent with previous works\cite{PhysRevB.99.195112,PhysRevB.99.224418,PhysRevB.103.195102,PhysRevB.105.094432,PhysRevLett.98.087204}.  
However, we also go beyond them, providing a more detailed classification of phases.

The spin-$1/2$ Kitaev ladder can be solved analytically via Jordan-Wigner transformation, which takes a 1D spin-$1/2$ system to a (Majorana) fermionic system by a specific non-local transformation.
A further Fourier analysis gives the energy spectrum and predicts the energy gap to close at $||K_x| - |K_y|| = |K_z|$.  The gap closing corresponds to a second order phase transition as confirmed by computing derivatives of the energy.
Details on the Jordan-Wigner transformation and energy spectrum derivation can be found in Appendix~\ref{appendix:JW}.

\begin{figure}
    \centering
    \includegraphics[width=0.95\linewidth]{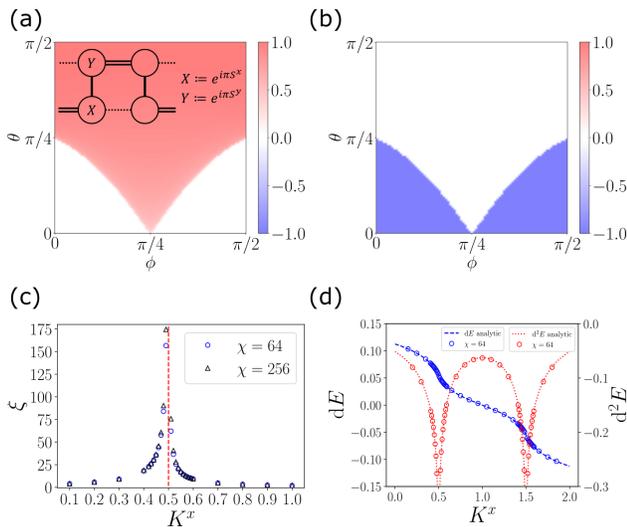}
    \caption{Results for spin-$1/2$ Kitaev ladder. The phase diagrams are found using a $4$-site iMPS with $\chi = 64$. (a) Order parameter $\expval{e^{i\pi S^x}e^{i\pi S^y}}$ measured on the sites of a $Z$-bond as shown in the inset, indicating SSB order. The axes show the parameter $K = (K_x, K_y, K_z)$ in spherical coordinates, i.e. $K_z = \sin\theta, K_x = \cos\theta\cos\phi, K_y = \cos\theta\sin\phi$.  (b) Pollmann-Turner order parameter of SPT order~\cite{pollmann2012detection}. (c) Correlation length on the line $K_x+K_y=2, K_z = 1$ to locate the critical point (d) Ground state energies and the 1st and 2nd derivatives, indicating second-order phase transitions.}
    \label{fig:spin1half}
\end{figure}

Although the analytical solution for spin-$1/2$ ladders gives efficient access to the ground state energy, numerical simulations with DMRG provide more efficient access to entanglement entropy, expectation values of operators such as the order parameter $\expval{\ox \oy}$ illustrated in Fig.~\ref{fig:spin1half}(a), and the detection of SPT order.
Additionally, numerical simulations help us compare the spin-$1/2$ Kitaev ladder with its higher-spin counterparts.

\paragraph*{Phase classification}  The gap-closings found analytically suggest that there are just three phases for $K_x$,$K_y$,$K_z>0$, each including one of the perturbative limits of large $K^z$, large $K^x$, and large $K^y$.
Using DMRG simulations directly in the thermodynamic limit, we confirm the presence of exactly three distinct phases, and we determine the nature of the phases and the transitions between them. 
In particular, we look for an SSB phase corresponding to the half-integer $Z$-limit perturbation theory by numerically evaluating the order parameter $\expval{\tau^y} = \expval{e^{i\pi S^x}e^{i\pi S^y}}$ and for SPT order protected by $\Sigma^Z_l \times \Sigma^Z_u$ using both the method of Pollmann and Turner~\cite{pollmann2012detection} (see also Appendix~\ref{appendix:SPT}) and string order parameters (SOPs).

Measurements of the order parameter $\expval{e^{i\pi S^x}e^{i\pi S^y}}$ show the existence of a large SSB phase. This phase starts from the large-$Z$ limit as we have predicted in the perturbation theory, with the expectation value of the order parameter very close to $1$ just like in the Ising model. The computed value of the SSB order parameter throughout parameter space is presented in Fig.~\ref{fig:spin1half}(a).

The two SPT phases are found around the large-$K_x$ and large-$K_y$ limits, confirmed by directly extracting the projective representation of symmetries from the MPS.
The MPS state is an SPT if the projective measurement is $-1$, and is a trivial symmetric state if it is $+1$. The symmetries of interest are broken if it is $0$. 
The results of the projective representation measurement are presented in Fig.~\ref{fig:spin1half}(b).

This conventional technique to detect SPT order, however, fails to distinguish between the two SPT phases in the half-integer spin Kitaev ladder, as it maps the two-leg ladders into 1D chains, compressing some information and losing the distinction between SPT-x and SPT-y.  We now show that they are, in fact, distinct phases.

Considering just the effective cluster model derived in the $X$- and $Y$-limits using perturbation theory, $H \propto \tau^z\tau^x\tau^z$, the protecting symmetry group is $\mathbb{Z}_2\times \mathbb{Z}_2$ and the corresponding SPT classification is $\mathbb{Z}_2$; in other words, there is one SPT phase and one trivial phase.~\cite{verresen2017one}  Then how can there be two different SPT phases, in addition to a trivial phase?

The answer lies in the mapping from the original spin ladder to the effective cluster model, as discussed in Sec.~\ref{sec:half_X_lim}.  Although the effective model is the same in each case, the protecting $\mathbb{Z}_2\times \mathbb{Z}_2$ symmetry groups are not.  Or, more precisely, some pre-images of $\mathbb{Z}_2\times \mathbb{Z}_2$ in the $X$-limit are different from some pre-images in the $Y$-limit, and we can show using SOPs that these different symmetry groups do indeed protect the respective SPT phases.

\begin{figure}
    \centering
    \includegraphics[width=0.98\linewidth]{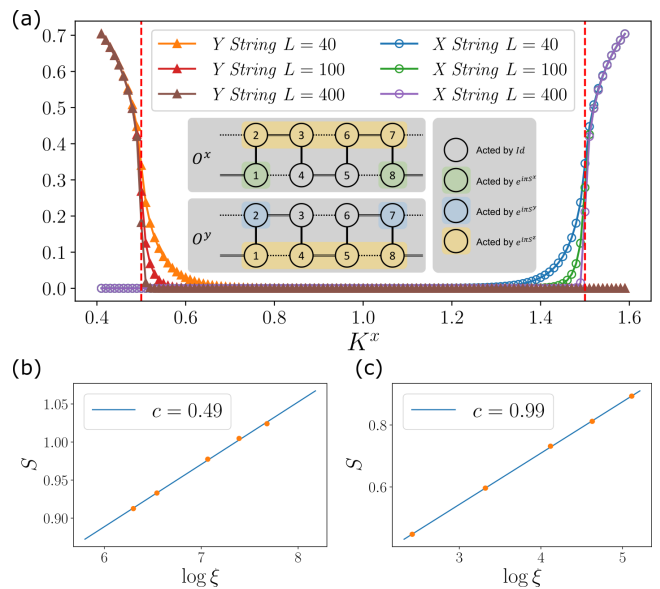}
    \caption{(a) Distinction between two SPT orders along the line $K_x + K_y = 2, K_z = 1$. The $x$-axis labels the value of $K^x$ and $K^y = 2-K^x$. In the inset we depict the distinction between the two SOPs for the two SPT phases. (b) Using finite-entanglement scaling to estimate the central charge $c=0.5$ of the transition point at $K = (1.5, 0.5, 1.0)$. (c) Using finite-entanglement scaling to estimate the central charge $c=1.0$ of the transition point at $K = (1.0, 1.0, 0)$. }
    \label{fig:string}
\end{figure}

In particular, one candidate $\mathbb{Z}_2\times \mathbb{Z}_2$ symmetry group that could protect the SPT-$x$ phase is $\Sigma^X_u \Sigma^Y_l \cross \Sigma^Z_u$, with a corresponding SOP $ O_x = e^{i\pi S^x_1} \left( \prod_{n=1}^{N} e^{i\pi S^z_{4n-2}}e^{i\pi S^z_{4n-1}} \right) e^{i\pi S^x_{4N }} $, illustrated in the inset of Fig.~\ref{fig:string}.  As shown in the figure, this SOP identifies SPT-$x$ as an SPT phase and SPT-$y$ as trivial.  On the other hand, a corresponding candidate $\mathbb{Z}_2\times \mathbb{Z}_2$ group that could protect SPT-$y$ is $\Sigma^X_u \Sigma^Y_l \cross \Sigma^Z_l$, with a corresponding SOP $ O_y = e^{i\pi S^y_2} \left( \prod_{n=1}^{N} e^{i\pi S^z_{4n-3}}e^{i\pi S^z_{4n}} \right) e^{i\pi S^y_{4N-1}} $.  As we again show in Fig.~\ref{fig:string}, this SOP identifies the SPT-$y$ phase as an SPT and the SPT-$x$ phase as trivial.  We provide further insights into the construction process of such SOPs, in particular into ambiguity in defining them due to the local symmetries of the Hamiltonian, in Appendix~\ref{appendix:SOPs}. 

As an additional perspective, note that the local Hilbert space here is spin-$1/2$, which is in the projective representation of any global symmetry; notably, on a single site the two generators of the protecting symmetry group $\Sigma^X_u \Sigma^Y_l \cross \Sigma^Z_u$ anti-commute.
To relate to the physics of a known SPT phase, two spin-$1/2$ must be combined into a linear representation, also guaranteeing that the symmetry group generators commute.
Specifically, spins on the $x$-bond ($y$-bond) can be combined into $X$-dimers ($Y$-dimers), transforming SPT-$x$ (SPT-$y$) into a topological phase while SPT-$y$ (SPT-$x$) becomes topologically trivial.

In the phase diagram, there is a single point $K_z=0, K_x=K_y$ where the two phases meet, and two simultaneous phase transitions occur: $X$ dimers transition from trivial to SPT, and $Y$ dimers transition from SPT to trivial. This could be interpreted as two co-existing $c=0.5$ gapless phase transitions, together resulting in the observed $c=1$ gapless phase transition [see Fig.~\ref{fig:string}].

\paragraph*{Phase transitions} The exact solvability of the spin-$1/2$ Kitaev ladder allows us to determine the nature of the phase transitions by studying the ground state energy and its derivatives as a function of ${\bf K}$. We show the analytically computed first and second derivatives across the SSB-SPT transition in Fig.~\ref{fig:spin1half} (d); we also show in the same figure the energy derivatives computed from DMRG, which agree precisely with the exact results, thus confirming the accuracy of our numerical simulations.

The SSB-SPT transition is evidently second-order, as we see both from the discontinuity of second-order energy derivatives and from the diverging correlation length (Fig.~\ref{fig:spin1half}(c)).
The nature of the transition can be determined more precisely by fitting the central charge according to the finite-scaling formula $S=\frac{c}{6} \log \xi + a$~\cite{pollmann2009theory}. As shown in Fig.~\ref{fig:string} (b) we find $c=0.5$. 

As discussed before, the transition between SPT-$x$ and SPT-$y$ is also continuous, but  $c=1$ since it should be regarded as a superposition of two $c=0.5$ transition.

\subsection{Phase diagram of spin-1 Kitaev ladder\label{sec:results_spin1}} 
\begin{figure}
    \centering
    \includegraphics[width=1.\linewidth]{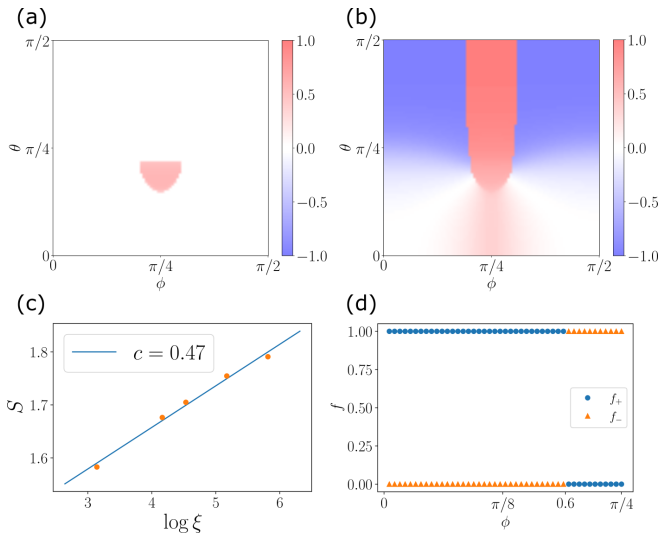}
    \caption{Phase Diagram of Spin-$1$ Kitaev Ladder. The phase diagrams are found using a $4$-site iMPS with $\chi = 64$. (a) SSB phase indicated by the order parameter $\expval{S^x S^y}$. (b) The value of $\expval{\ox\oy}$ in the phase diagram. (c) Using finite-entanglement scaling to estimate the central charge of the phase transition point near $K=(1.4, 1.4, 1.0)$ between the SSB and trivial phases. (d) Overlaps between GS and two different dimer product states change suddenly at the first order phase transition around $\phi = 0.6$ with $\tan\phi=0.69$ as predicted. } 
    \label{fig:spin1pd}
\end{figure}

For spin-$1$ Kitaev ladders, we find a very different phase diagram.  
As shown in Fig.~\ref{fig:spin1pd}, there are only SSB and trivial paramagnetic phases.
These numerical results are consistent with the prediction of perturbation theory, namely that there are only trivial phases in the anisotropic limits.

\paragraph*{Phase classification} Recall from Eq.~\eqref{eq:unnecessaryPT} we have concluded that the effective Hamiltonian in the large-$Z$ limit is given by the combination of $\tau^x_i$ and $\tau^x_i \tau^x_{i+1}$ along with coefficients $\alpha$ and $\beta$, respectively; by explicit calculation of all contributions up to 4th order terms, we find that the coefficients for spin-$1$ are $\alpha = - (K_x^4 - \frac{18}{7} K_x^2 K_y^2 + K_y^4)/K_z^3$ and $\beta = - 5  K_x^2 K_y^2/K_z^3$.
This predicts a first-order transition between different product states, $\ket{+}_x^{\otimes N}\leftrightarrow\ket{-}_x^{\otimes N}$, where $\ket{+}_x$ and $\ket{-}_x$ are the $\pm 1$ eigenstates of $\tau^x$, respectively.  The transition occurs when $\alpha(K)$ changes sign, at $\tan\phi=K_x / K_y = \sqrt{\frac{1}{7}(9\pm 4 \sqrt{2})}=1.44$ and $0.69$.
Converting $\tau^x$ back to the original model, we get the operator $e^{i \pi S^x_0} e^{i \pi S^x_1}$ or equivalently $-e^{i \pi S^x_0} e^{i \pi S^y_1}$; we show the expectation value of the latter operator in Fig.~\ref{fig:spin1pd}(b), and the first order transition is immediately apparent in the $Z$-limit, at $\theta\approx\frac{\pi}{2}$.~\footnote{Note that, although $\langle \tau^x\rangle = \left\langle-e^{i \pi S^x_0} e^{i \pi S^y_1}\right\rangle$ shows the unnecessary phase transition, it commutes with all global symmetries and is thus not an order parameter.} Alternatively, we can observe the transition by computing the overlap, $f_\pm = \abs{\bra{\psi}\ket{d_{z, \pm}}}$, of the MPS ground state from DMRG with two dimer states
\begin{equation}
    \ket{d_{z,\pm}} = \otimes_{\expval{ij}_z} \frac{1}{2}\left(\ket{+S}_i\ket{-S}_j \pm \ket{-S}_i\ket{+S}_j\right),
\end{equation}
which are precisely the product states $\ket{+}_x^{\otimes N}$ and $\ket{-}_x^{\otimes N}$ in the effective model from perturbation theory.  The overlaps are shown in Fig.~\ref{fig:spin1pd}(d).

Unlike in the case of spin-1/2, for spin-1 we also find a phase  that is not present in the effective models from perturbation theory.  As shown in Fig.~\ref{fig:spin1pd}(a), the order parameter $\langle S^x_0 S^y_1 \rangle$ reveals an SSB phase around the isotropic point.  We note that, while this is the same operator used to detect SSB in the spin-1/2 case, there it also corresponded to $\tau^y$ in the perturbative model in the $Z$-limit, whereas here it becomes the zero operator in any anisotropic limit.

\paragraph*{Phase transitions} The unnecessary phase transition between trivial phases is first-order, as verified by directly measuring the expectation value of $\tau^x$ as in Fig.~\ref{fig:spin1pd}.

In contrast, the critical points between SSB and trivial phases are second order and Ising-like, with $c = 0.5$.

\subsection{Phase diagram of spin-3/2 Kitaev ladder} 
\paragraph*{Phase classification} As predicted in the perturbation theory for arbitrary half-integer spin-$S$ Kitaev ladder, the phase diagram for spin-$3/2$ is quite similar to that of spin-$1/2$: we again have SSB, SPT-$x$, and SPT-$y$ phases.

\begin{figure}
    \centering
    \includegraphics[width=1.\linewidth]{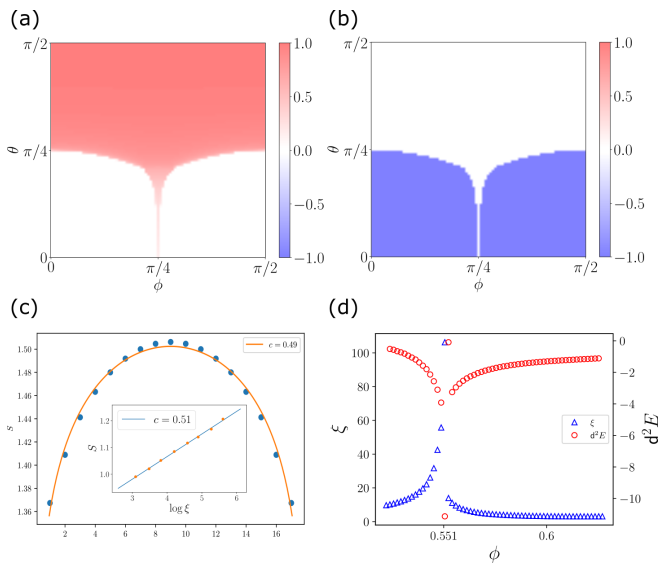}
    \caption{Phase diagram of spin-$3/2$ ladders. The phase diagrams are given by a $4$-site iMPS with $\chi = 128$. (a) Order parameter $\expval{\ox\oy}$. (b) Pollmann-Turner order parameter for SPT order~\cite{pollmann2012detection}. (c) Central charge estimation at a critical point $\theta = \frac{2}{9}\pi, \phi = 0.55$, giving $c\approx 0.5$. (d) Correlation length $\xi$ and second derivative of energy $d^2E/d^2\phi$ along $\theta = \frac{2}{9} \pi$.}
    \label{fig:spin3half_pd}
\end{figure}

To detect the SSB order, we again use the order parameter $\langle\tau^y\rangle$. 
For spin-$1/2$ this order operator happens to be the simple $S^x S^y$ but for spin-$3/2$ it is $e^{i\pi S^x}e^{i\pi S^y}$.   
Order parameter measurements are summarized in Fig.~\ref{fig:spin3half_pd} (a).

The SPT phases look the same in the perturbative limits for all half-integer spin-$S$ cases.
Thus projective representations are  detected by exactly the same technique on the same set of symmetries as in the spin-$1/2$ case. 
SPT order measurement results are summarized in Fig.~\ref{fig:spin3half_pd} (b).

\paragraph*{Phase transitions} In comparison with the spin-$1/2$ phase diagram, now the critical lines between SSB and SPT sit much closer to each other and the SSB phase is smaller when approaching the point $K^z=0, K^x=K^y$.
From Fig.~\ref{fig:spin3half_pd} (d) we can see that the correlation length peaks near the critical point, and that there is a discontinuity of the second-order energy derivative as for spin-$1/2$.
Together, these show that the transition is again second order.
As shown in Fig.~\ref{fig:spin3half_pd} (c) both iDMRG and fDMRG give the same fitting of central charge $c=0.5$.

\subsection{Phase diagram of spin-2 Kitaev ladder\label{sec:results_spin2}}
\begin{figure}
    \centering
    \includegraphics[width=1.\linewidth]{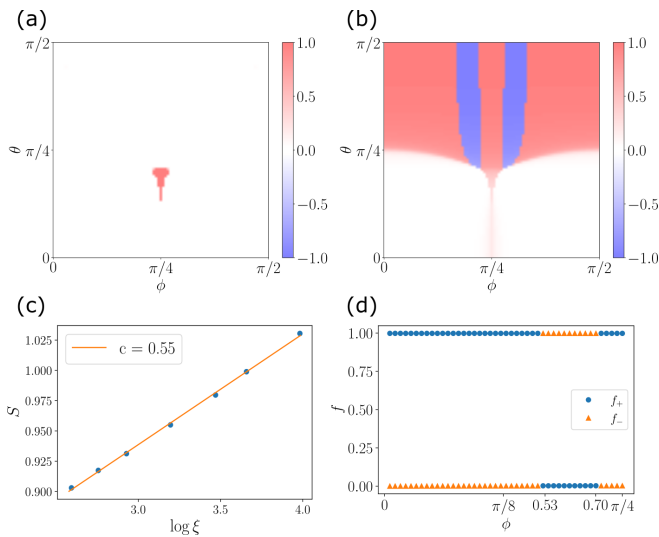}
    \caption{Phase Diagram of Spin-$2$ Kitaev Ladder. The phase diagrams are given by a $4$-site iMPS with $\chi = 128$. (a) The phase diagram given by the order parameter $\expval{S^x S^y}$. (b) The value of $\expval{\ox\oy}$ in the phase diagram. (c) Central charge estimation at a transition point. (d) Overlaps between GS and two different dimer product states change suddenly at the first order phase transition points around $\phi = 0.53$ and $\phi = 0.70$.}
    \label{fig:spin2}
\end{figure}

Similarly to the case of spin-1, there are different trivial phases smoothly connected to product states of dimers, again with first-order unnecessary phase transitions.
Also, the same SSB phase as in the spin-$1$ case exists in the central region, although the phase is much smaller for spin-2.
The computed SSB order parameter and observable indicating the unnecessary phase transitions are shown in Figs.~\ref{fig:spin2}(a) and (b), respectively.  The unnecessary phase transitions are also observable from the overlap with different dimer product states, shown in Fig.~\ref{fig:spin2}(d).

Additionally, the transition between the trivial phases and the SSB phase again has central charge $c=0.5$, as in the case of spin-1.
The finite-entanglement scaling to determine $c$ is shown in Fig.~\ref{fig:spin2}(c).

We note the interesting fact that Fig.~\ref{fig:spin3half_pd}(a) and Fig.~\ref{fig:spin2}(b) are quite visually similar, even though in the former the dividing line between red and white regions shows a distinct phase transition between SSB and SPT order and in the latter it shows a smooth (albeit steep) change in an expectation value within a single phase.  The apparent correspondence may be due to the expected convergence of integer and half-integer behavior in the classical large-$S$ limit.

\section{Discussion\label{sec:discuss}}
In this work, we have analyzed the general behaviors of higher-spin Kitaev ladders  and concluded that for half-integer spins they have phases with SSB and SPT order with an Ising-like transition between them, and that for integer spins there are SSB and trivial phases in the phase diagram.
The phases in the half-integer case can be understood through perturbation theory, as can the trivial phase in the integer case.
Numerical results from analytical solutions and DMRG simulations on finite and infinite ladders confirm the results of perturbation theory, find the SSB phase for integer spins, and identify further detailed properties of the phase diagrams.
With this combination of different theoretical and numerical methods, we provide a thorough and conclusive analysis on general features of spin-$S$ Kitaev ladders.

For several values of $S$, we investigate the transitions between different phases by finite-entanglement scaling, and we identify both first-order and second-order phase transitions: the transitions between trivial phases are first-order and the transitions between SSB and SPT/trivial are second-order.
Most of the second-order phase transition points are Ising-like with central charge $c=0.5$. The exception is the transition between distinct SPT phases, distinguished by SOPs, which has $c=1$. 

Finally, we turn to the question of what our results imply for the full 2D Kitaev honeycomb model with higher spin.  We work by analogy: we relate the spin-$1/2$ ladder and spin-$1/2$ honeycomb, and we hypothesize an analogous relationship between higher-spin ladders and higher-spin honeycombs, allowing us to predict the behavior of the latter from the former.

We first note that perturbation theory in the anisotropic limits of the spin-1/2 Kitaev honeycomb gives the toric code model, as shown by Kitaev in the original paper~\cite{KITAEV20062}, and recently demonstrated on the spin-$S$ honeycomb was given recently~\cite{Minakawa2019PT}.
In fact, the toric code is the 2D analogue of the cluster model we found in the half-integer-spin ladders in the $X$-limit, thus suggesting that the behavior in the $X$-limit of our model carries over to 2D while the $Z$-limit behavior does not.  We find that on the ladder, the spin-3/2 case exhibits the same cluster model SPT phases as in spin-1/2, and in fact they are stabilized by the higher spin, covering a larger portion of the phase diagram.  We thus surmize that on the full honeycomb, phases with gapped $\mathbb{Z}_2$ topological order will be present in the anisotropic limits for higher half-integer spin just as for spin-1/2, and indeed they will reach farther towards the isotropic point.
On the other hand, since the SSB phase appears in the spin-1/2 ladder but does not appear at all in the spin-1/2 honeycomb model, there is no good reason to think it would appear in higher-spin honeycomb models either.
We cannot make a clear prediction for the isotropic point itself, as there is no clear analogue for the ladder.  

For integer spin, if we again assume that the $X$-limit of the ladder is analogous to the anisotropic limits of the honeycomb, we can predict trivial phases in all anisotropic limits, smoothly connected to product states of dimers along the corresponding strongest bonds.  Unlike for half-integer spin, we find an  SSB phase distinct from any anisotropic limit in a small region around the isotropic point on both spin-1 and spin-2 ladders. 
We also confirmed that the SSB at the isotropic point survives in the large-S limit by computing up to spin-4 and observing convergence of the SSB order parameter with increasing spin. 
For more details, please check the Appendix~\ref{appendix:SSB}.
It is not clear whether or not this SSB phase would carry over to the integer-spin honeycomb models.


\begin{acknowledgments}

Thanks to Ruochen Ma, Yijian Zou and Weicheng Ye for multiple helpful discussions.
Research at Perimeter Institute is supported in part by the Government of Canada through the Department of Innovation, Science and Economic Development Canada and by the Province of Ontario through the Ministry of Colleges and Universities.
Y.C. acknowledges  supports  from  the  Natural Sciences and Engineering Research Council of Canada(NSERC)  through  Discovery  Grants.
This research was enabled in part by support provided by WestGrid and the Digital Research Alliance of Canada\cite{alliancecan2023}.
Part of the computation was done on the Symmetry cluster of Perimeter Institute.
\end{acknowledgments}


\bibliography{references}
\appendix
\include{appendix}
\end{document}

%% file: appendix.tex
\newpage 
\newcommand{\HGS}{\mathcal{H}_{\text{GS}}}
\newcommand{\HES}{\mathcal{H}_{\text{ES}}}
\newcommand{\PGS}{\mathcal{P}_{\text{GS}}}
\newcommand{\PES}{\tilde{\mathcal{P}}_{\text{ES}}}
\section{\texorpdfstring{Analytic solution to spin-$1/2$ Kitaev ladder}{Analytic solution to spin-1/2 Kitaev ladder}\label{appendix:JW}}
To analytically solve the Kitaev ladder, we need to express the original model in a 1D order.
In the `snake' ordering, the original Kitaev ladder with $4N$ spins is mapped to a 1D chain with the bulk Hamiltonian
\begin{equation}
    H(K) = \sum_{n = 1}^{2N} K_z {\sigma}^z_{2n - 1} {\sigma}^z_{2n} + K_x {\sigma}^x_{2n} {\sigma}^x_{2n + 1} + K_y {\sigma}^y_{2n - 1} {\sigma}^y_{2n + 2}
\end{equation}
where we denote the operators as $\sigma$ to emphasize the difference from higher spin-$S$ counterparts.
Such a snake ordering exploits the translation symmetry $\mathbb{T}_2$ over two sites instead of four sites.
For periodic boundary conditions (PBCs) we can take $\forall m>0, {\sigma}_{4N+m} = {\sigma}_{m}$ and for open boundary conditions (OBCs)
${\sigma}_{4N+m} = 0$.
Therefore, the difference between OBC and PBC lies in the fact that a PBC Hamiltonian has two more terms at the boundary: $K_x \sigma^x_{4N} \sigma^x_{1}$ and $ K_y\sigma^y_{4N-1} \sigma^y_{2}$.

\subsection{Jordan-Wigner transformation}
\begin{figure}
    \centering
    \includegraphics[width=0.95\linewidth]{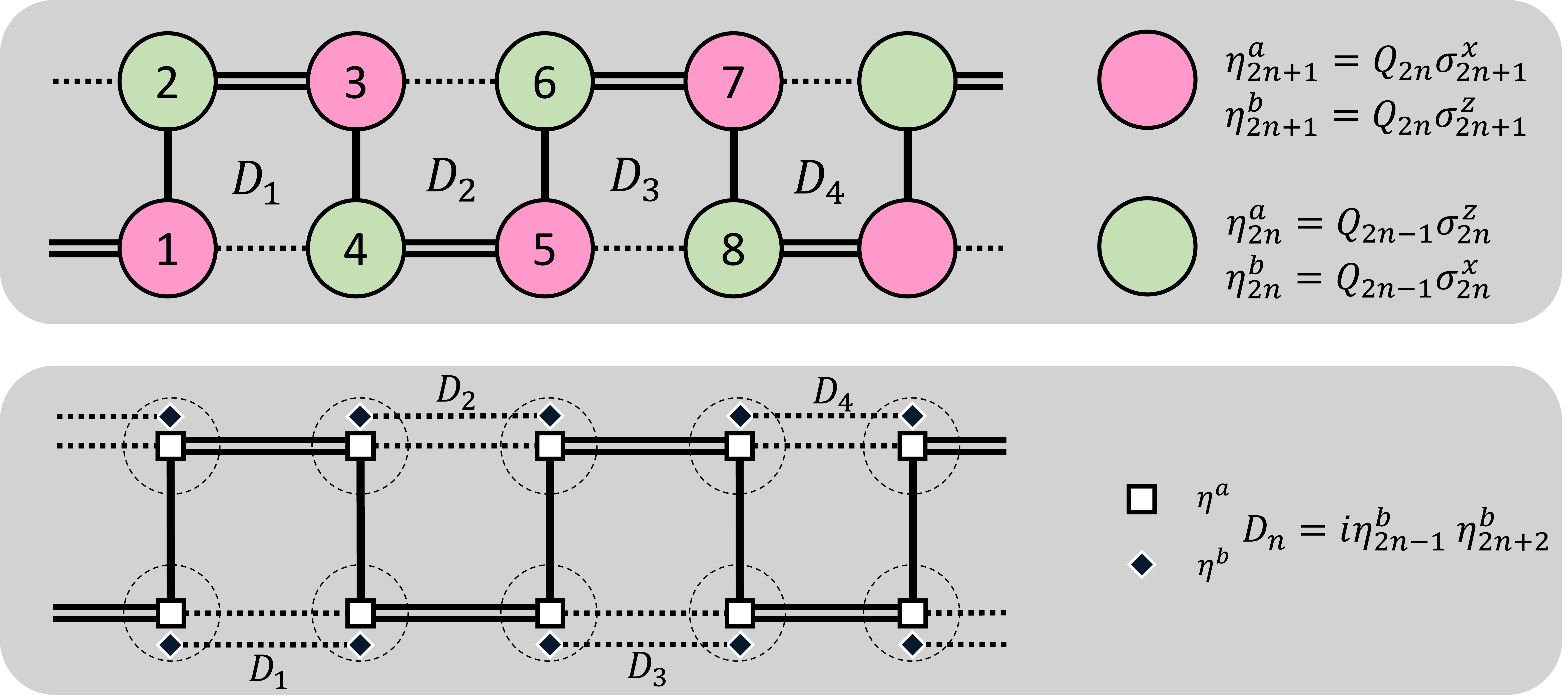}
    \caption{Jordan-Wigner transformation for spin-$1/2$ Kitaev Ladder. The upper panel represents the original spin model with the definition of the JWT, and the lower panel shows the Hamiltonian in the Majorana representation. Note that in this representation, the specific transformation we chose makes the local symmetries $D_n$ live only on the bonds between $K_y$ pairs.}
    \label{fig:JWT_Majorana}
\end{figure}

Jordan-Wigner transformation (JWT) is a frequently used technique for analytically solving some 1D spin-$1/2$ chains by nonlocally mapping the Pauli matrices to fermionic operators (see Fig.~\ref{fig:JWT_Majorana}).
Here we use the following transformation:
\begin{equation}
    \begin{aligned}
    \eta^a_{2n} &= Q_{2n - 1} \sigma^z_{2n} \\
    \eta^b_{2n} &= Q_{2n - 1} \sigma^x_{2n} \\
    \eta^a_{2n+1} &= Q_{2n} \sigma^x_{2n+1} \\
    \eta^b_{2n+1} &= Q_{2n} \sigma^z_{2n+1} \\
    \end{aligned}
\end{equation}
where $Q_n := \prod_{k=1}^n \sigma^y_k, Q_0=1$ is a string operator.
The anti-commutation relations between the Majorana fermions are also straightforward, $\{\eta^\alpha_i, \eta^\beta_j\}=2\delta^{\alpha, \beta} \delta_{i,j}$.

In this way, the three types of interactions in the snake-order Kitaev ladder can be reformulated as 
\begin{equation}\label{eq:JWTinter}
    \begin{aligned}
    \sigma^x_{2n} \sigma^x_{2n+1} &= i \eta^a_{2n} \eta^a_{2n+1} \\
    \sigma^z_{2n-1} \sigma^z_{2n} &= -i \eta^a_{2n-1} \eta^a_{2n} \\
    \sigma^y_{2n-1} \sigma^y_{2n+2} &=  (i \eta^a_{2n-1} \eta^a_{2n+2} )(i \eta^b_{2n-1} \eta^b_{2n+2} )\\
    \end{aligned}
\end{equation}

\noindent After JWT, the bulk Hamiltonian is
\begin{equation}\label{eq:JWTH}
\begin{aligned}
    H(K) &= -K_z \sum_{n = 1}^{2N} \left. i \eta^a_{2n-1} \eta^a_{2n} \right.\\ 
    &\quad + K_x \sum_{n = 1}^{2N-1}i \eta^a_{2n} \eta^a_{2n+1} \\ 
    &\left. \quad + K_y \sum_{n = 1}^{2N-1}(i \eta^a_{2n-1} \eta^a_{2n+2} )(i \eta^b_{2n-1} \eta^b_{2n+2} ) \right.\\
\end{aligned}
\end{equation}
where apparently there are many terms $(i \eta^b_{2n-1} \eta^b_{2n+2} )$ commuting with the Hamiltonian, which turn out to be just the local symmetries 
\begin{equation}
D_n=(i \eta^b_{2n-1} \eta^b_{2n+2} ) = \sigma^x_{2n-1} \sigma^y_{2n} \sigma^y_{2n+1} \sigma^x_{2n+2}
\end{equation}
as defined in Sec.~\ref{subsection: symmetries}. Since each $D_n$ is the product of two Majorana fermions, each of them controls a subspace of dimension 2.  Intuitively, the local symmetries can be viewed as ``$\mathbb{Z}_2$-fluxes" attached to pairs of Majorana fermions.

JWT usually introduces many `cluster' terms in the Majorana representation for generic spin chains, but here fortunately the two-body interactions get transformed to two-body interactions.  
The only exceptions are the $\sigma^y \sigma^y$ terms, but they also have a convenient form: two-body interactions multiplied by local symmetries.

Since the local symmetries commute with the Hamiltonian and each other, we can divide the Hilbert space into a direct sum of subspaces, each a simultaneous eigenspace of all the $D_n$. Within each such subspace, specified by the list of eigenvalues $(D_1, D_2, \cdots, D_{2N}), D_n\in\pm1$, each $D_n$ can be treated as a constant in Eq.~\eqref{eq:JWTH}.  Thus the Hamiltonian within each block is effectively quadratic and can be solved efficiently.

\subsection{Thermodynamic limit}
Assuming the states of interest live in a subspace such that $D_n=D=\pm1$, i.e. the flux configuration is translationally invariant, the fermionic Hamiltonian can not only be solved efficiently as a single-body problem, but in fact can be solved analytically by Fourier transform. The Hamiltonian in momentum space is 
\begin{equation}
    H = i\sum_k \alpha_k \beta_k^\dagger \left( K_z + K_x e^{ik}+ (DK_y)e^{-ik} \right)
\end{equation}
where $\eta^a_{2n - 1} = \sum_{k} e^{i n k} \alpha_k$ and $\eta^a_{2n} = \sum_{k} e^{i n k} \beta_k, k \in \{ \frac{\pi (2m+1)}{2 N}|-N, -N+1, \cdots, N-1 \}$.
The corresponding spectrum
\begin{equation}
    \epsilon(k;\mathbf{K}, D) = \sqrt{(K_z + P_+\cos k)^2 + P_-^2\sin^2k}
\end{equation}
where $P_\pm = K_x \pm DK_y$.

Integrating over the momentum space gives the energy per site and $D=\pm1$ give different energies
\begin{equation}
    E(\mathbf{K}, D) = -\int_0^{2\pi} \dd k \epsilon(k;\mathbf{K}, D).
\end{equation} 
This is the energy function we used in Fig.~\ref{fig:spin1half} (d), where the critical points are evident from the singularity of $\frac{\dd^2 E}{\dd\lambda^2}$, with $\lambda$ the linear parameter controlling the change on a specific curve in parameter space for $\mathbf{K}$.

\subsection{Finite size: OBC and PBC}
For finite-size ladders with either OBC or PBC, with a fixed choice of the $\{D_n\}$, as noted above we can analyze the system by efficiently solving a quadratic (single-body) Hamiltonian.

The procedure is as follows: let $L=2N$ be the number of two-site unit cells where $2L = 4N$ is the total number of spins, along with a fixed flux configuration $D=(D_n), n=1,\cdots,L$, the Hamiltonian can be rewritten as $H=i\frac{1}{2}\eta^\dagger M \eta$ where $\eta^\dagger :=(\eta^a_1, \cdots, \eta^a_{4N})$ and $M$ is a skew-symmetric matrix depending on $K_x, K_y, K_z$ and $D$.
Schur decomposition over $M$ gives a transformation $M = O^T N O$ where $N = \bigoplus_{n=1}^{L} \epsilon_n \begin{bmatrix}
0 & 1 \\
-1 & 0 
\end{bmatrix}, \epsilon_n \le 0$ is the spectrum and $O$ is the corresponding orthogonal matrix linearly combining the Majorana fermions. 
Then $H = i\frac{1}{2}\zeta^\dagger N \zeta$ where $\zeta = O \eta$ and eventually the Hamiltonian is decomposed as $H=\sum_{n=1}^{L} i\epsilon_n \zeta_{2n-1} \zeta_{2n}$.

Recalling the JWT relationship between spin two-body interactions and Majorana two-body interactions in Eq.~\eqref{eq:JWTinter}, we can see that now the whole system can be transformed to a new spin-1/2 system (denoted by $\rho$) with only $L$ non-interacting dimers, $H = \sum_{n=1}^{L}\epsilon_n \rho^z_{2n-1} \rho^z_{2n}$.
Note that these spins are different from those of the original model; the total number of spins is reduced from $2L$ to $L$ since we fixed the $D_n$.

This method can be used to calculate the exact ground state energy of finite spin chains by simply summing over the spectrum $E = -\sum_{n=1}^{L}\epsilon_n$.
Note that, starting from the finite OBC/PBC calculation, some analysis on the appearance of zero modes $\epsilon_n=0$ with $L \rightarrow \infty$ gives the degeneracy of the original systems in the thermodynamic limit.

\section{Perturbation theory}\label{appendix:PT}

\subsection{Brief review of perturbation theory}
In perturbation theory, we divide the full Hamiltonian into the unperturbed Hamiltonian and perturbing Hamiltonian, i.e.
\begin{equation}
    H = H_0 + H^\prime
\end{equation}
where $H_0$ is usually simple, so that its ground subspace (or even all the eigen-subspaces, which is the case for Kitaev ladders) is easy to identify.  In this way, the full Hilbert space $\mathcal{H}$ is also partitioned into $\mathcal{H} = \mathcal{H}_{\text{GS}} \bigoplus \mathcal{H}_{\text{ES}}$ determined by $H_0$, and there is no component of $H_0$ connecting $\HGS$ and $\HES$ (i.e. off-diagonal components of $H_0$ are zero).
Usually $\HGS$ is highly degenerate, e.g. $2^{N/2}$-dimensional in $N$-spin-1/2 Kitaev ladders.  Our goal is to find how $H^\prime$ helps split the degeneracy of $\HGS$ and thus to find the effective Hamiltonian of $H$ projected to $\HGS$.

In particular, the degeneracy is split by `virtual' (or `off-shell') processes in which repeated applications of $H^\prime$ `pump out' states from $\HGS$ to $\HES$ and eventually back to $\HGS$.  In other words, second and higher orders of $H^\prime$ can produce non-trivial matrix elements within $\HGS$, creating nontrivial dynamics and lifting the degeneracy.

The resulting effective Hamiltonian is 
\begin{equation}
    H_{\text{eff}} = E_0+ \sum_{n=1}^\infty \PGS(H^\prime\PES)^{n-1} H^\prime \PGS
\end{equation}
where the projection operators are $\PGS = \sum_{\psi\in\HGS}\ket{\psi}\bra{\psi}$, $\PES = \sum_{\psi\in\HES}\frac{1}{E_0 - E_\psi}\ket{\psi}\bra{\psi}$.

\subsection{Application to Kitaev ladders}

As discussed in Sec.~\ref{sec:Sec2}, the ground subspace $\mathcal{H}_{\text{GS}}$ and the perturbing Hamiltonian $H^\prime$ are determined by the limit we consider.  In the limit of large $K_x$, $\HGS$ is effectively a chain of spin-$1/2$ sites corresponding to dimers on the $S^xS^x$ bonds, and $H^\prime$ is given by the summation over $S^yS^y$ and $S^zS^z$.  Correspondingly, in the large $K_z$ limit, dimers are on the $ZZ$ bonds and the perturbing Hamiltonian comes from $XX$ and $YY$.

To change the state of one dimer from $\ket{0}=\ket{+S}\ket{-S}$ to $\ket{1}=\ket{-S}\ket{+S}$, each of the two spins must be raised or lowered $2S$ times.
Recalling the definition for the ladder operators of spins $S^\pm = S^x \pm i S^y$, the perturbing terms can implement the transition between distinct states in $\mathcal{H}_{\text{GS}}$ with a minimum order $4S$.

We now explicitly describe how to carry out the perturbation theory in each of the four cases: $X$- and $Z$-limits with integer and half-integer spin.
Recall the Hamiltonian for Kitaev ladders to find the fundamental difference bewteen $Z$-limit and $X$-limit: the former maps to a spin-$1/2$ chain where each spin has two neighbors and the latter to a chain where each spin has four neighbors.

\begin{figure}
    \centering
    \includegraphics[width=0.85\linewidth]{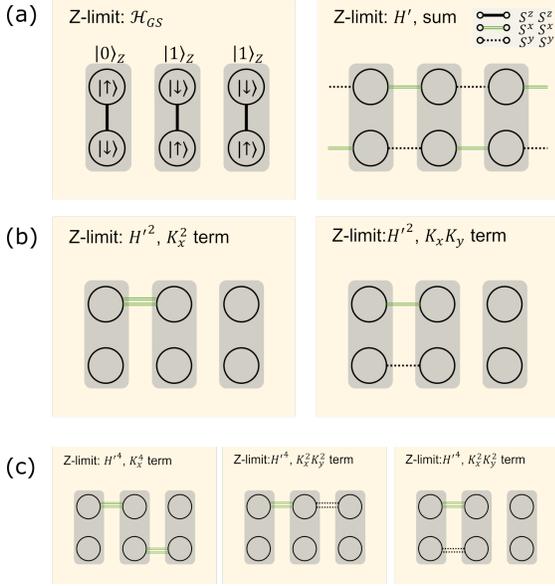}
    \caption{Diagrams that contribute to the perturbation theory of Kitaev ladders in the  $Z$-limit}
    \label{fig:PTzlim}
\end{figure}

\paragraph{Half-Integer $S$ in $Z$-limit} In this case, perturbation terms of the lowest order ($4S$, e.g. second order for spin-$1/2$) involve two connecting dimers. 
Diagrams showing terms that may contribute are listed in Fig.~\ref{fig:PTzlim}.

For the simplest example, we can explicitly write down the two-dimer Hamiltonian for spin-1/2 as $H=\sigma^z_1\otimes\sigma^z_2 + \sigma^z_3\otimes\sigma^z_4 + K_x \sigma^x_1\otimes\sigma^x_3 + K_y \sigma^y_2\otimes\sigma^y_4$, $0<K_x,K_y\ll 1$.
The lowest nonzero order of perturbation theory is $4S=2$; in Fig \ref{fig:PTzlim} (b) we present two diagrams showing representative second-order contributions.
The coefficients  $K_x^2$ and $K_x K_y$ show which terms in $H^\prime$ were applied for each diagram.
The resulting effective Hamiltonian is $\frac{1}{4}(K_x^2+K_y^2)\text{Id}+\frac{1}{2}K_x K_y \tau^y \otimes \tau^y$ with eigenvalues $\frac{1}{4}(K_x\pm K_y)^2$, each with a twofold degeneracy.
An exact diagonalization of $H$ indicates the true ground energy to be $-\sqrt{4+(K_x+K_y)^2} = -2 - \frac{1}{4}(K_x+K_y)^2 + O((K_x+K_y)^4)$; the $-2$ is the energy contribution from $H_0$.
For any higher half-integer spin, it is likewise the case that $4S$-order perturbations make non-trivial contributions.

\paragraph{Integer $S$ in $Z$-limit} In this case, the lowest order perturbation terms involve three connecting dimers. 
Still the lowest order of perturbation terms with nontrivial effective contribution is $4S$.

For example, for spin-$1$ Kitaev ladders, the 4th order terms with nonzero effect are presented in Fig \ref{fig:PTzlim} (c): the first one makes a $\tau^x$ contribution, the third makes both $\tau^x$ and $\tau^x \tau^x$ contributions, and the second only brings effective identity operators.
Therefore, we can guess that there may be a competition between $\tau^x$ and $\tau^x \tau^x$; in fact, they do not compete with each other since the final coefficient for $\tau^x \tau^x$ is negative and therefore only an effective ferromagnetic phase is preferred.
However, in the spin-$1$ case the coefficient of the $\tau^x$ term itself can take either sign depending on the precise values of $K_x$ and $K_y$ as specified in Sec.~\ref{sec:results_spin1}, leading to a first-order transition between different orientations of effective spins.  As shown in Fig.~\ref{fig:spin2}, a similar transition also exists for spin-$2$.  For higher-integer spin cases we expect the same behavior, but we have not checked explicitly.
\begin{figure}
    \centering
    \includegraphics[width=0.85\linewidth]{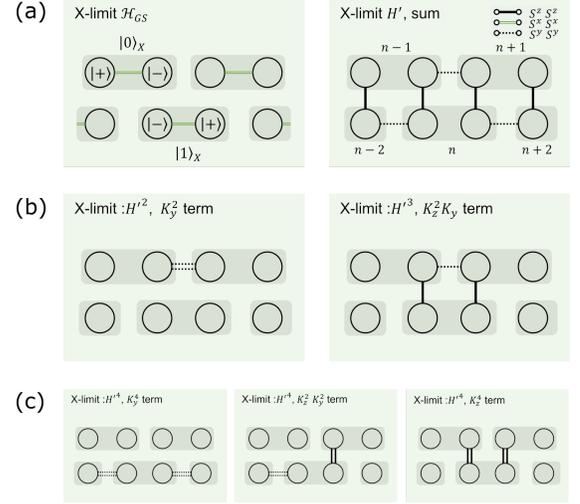}
    \caption{Diagrams that contribute to the perturbation theory of Kitaev ladders in the  $X$-limit}
    \label{fig:PTxlim}
\end{figure}

\paragraph{Half-Integer $S$ in $X$-limit} In this case, the lowest order perturbation terms involve three dimers. 
To see this, consider the right panel of Fig.~\ref{fig:PTxlim}(a). 
Here we label the five dimers that may be involved as in the right part of Fig~\ref{fig:PTxlim}(a) from $n-2$ to $n+2$.

First of all, unlike the $Z$-limit, the perturbation theory fails if we only consider two adjacent dimers, either the dimer pair $(n-1, n)$ or the pair $(n-1,n+1)$. 
The reason lies in the fact that they are only connected by one bond, of either $Z$-type or $Y$-type, which hits only one of the two spins in each dimer. Therefore, in each dimer, the other spin cannot be affected, so there is no chance to go back to $\HGS$ and all off-diagonal matrix elements in the effective Hamiltonian vanish.
The only non-zero contribution made by only considering two dimers comes from applying a single bond an even number of times (e.g. the left part of Fig~\ref{fig:PTxlim}(b)), which turns out to give a trivial energy shift for all states in $\HGS$.

On the other hand, a nonzero off-diagonal matrix element is possible when (at least) three dimers are included.
For example, for spin-$1/2$, the bonds shown on the right part of Fig.~\ref{fig:PTxlim}(b) flip both spins of dimer $n$ and the matrix element connecting $\ket{0}_{X,n}$ and $\ket{1}_{X,n}$ is non-zero; for an arbitrary $S$, we need $2S$ for each site and $4S$ in total.
Note that when $S$ is half-integer, $2S$ is an odd integer, so for dimers $n \pm 1$ one spin site is flipped an odd number of times and an extra bond connecting them is required.
Therefore, the lowest order of non-trivial terms in perturbation theory will be $4S+1$.
Such a bond configuration gives a term $\tau^z_{n-1} \tau^x_{n} \tau^z_{n+1}$ in the effective Hamiltonian; dimers $n \pm 1$ are not transformed as $\mathds{1}_{n\pm1}$ because one spin is acted on by an odd number ($2S$) of $S^z$ and an odd number ($1$) of $S^y$, causing a phase distinction between $\ket{+S}$ and $\ket{-S}$ on the spin sites of the original model, and hence also distinguishing between $\ket{0}$ and $\ket{1}$ in the effective model.

For spin-1/2, where $(4S+1)=3$, the diagram on the right of Fig.~\ref{fig:PTxlim}(b) is the only one allowed at order $(4S+1)$.  For higher half-integer spin, a term at order $(4S+1)$ could in principle involve more than three dimers.  One could imagine, for example, a seventh-order term that includes the three bonds from the right of Fig.~\ref{fig:PTxlim}(b) and the four bonds from one of the diagrams in Fig.~\ref{fig:PTxlim}(c).  However, such terms cannot act nontrivially on dimer $n$.  Ultimately, the only nontrivial term at order $(4S+1)$ for any half-integer spin is $\tau^z_{n-1} \tau^x_{n} \tau^z_{n+1}$ coming from the three-dimer configuration as on the right of Fig.~\ref{fig:PTxlim}(b) but with $2S$ vertical bonds on each side.

\paragraph{Integer $S$ in $X$-limit} In this case the lowest-order nontrivial perturbation terms require five connecting dimers. 
This case involves many different configurations, but all of them correspond to the same $\tau^x$ effective term.
Note that even though both the $Z$-limit and the $X$-limit for integer spins are effectively $\tau^x$, because $\HGS$ is different, $\tau^x$ in the two cases represents different `orientations' of the original model. 
Despite this distinction, the perturbative ground states are both trivial phases, $\ket{\Psi}_\gamma = \otimes_{\langle i,j\rangle_\gamma}\left( \ket{+S}_i\ket{-S}_j +  \ket{-S}_i\ket{+S}_j\right)$ where $\gamma = X, Z$. Indeed, they are part of the same phase since, as we show in Sections~\ref{sec:results_spin1} and \ref{sec:results_spin2}, they can be connected without crossing a phase boundary.

\section{SPT detection in MPS\label{appendix:SPT}}

The now standard technique for detecting SPT order in an MPS was given by Pollmann and Turner~\cite{pollmann2012detection}.
We review their method, then explain how it can be modified for the present case, where the SPT order occurs in the effective model in the reduced Hilbert space $\HGS$.

\begin{figure}
    \centering
    \includegraphics[width=0.95\linewidth]{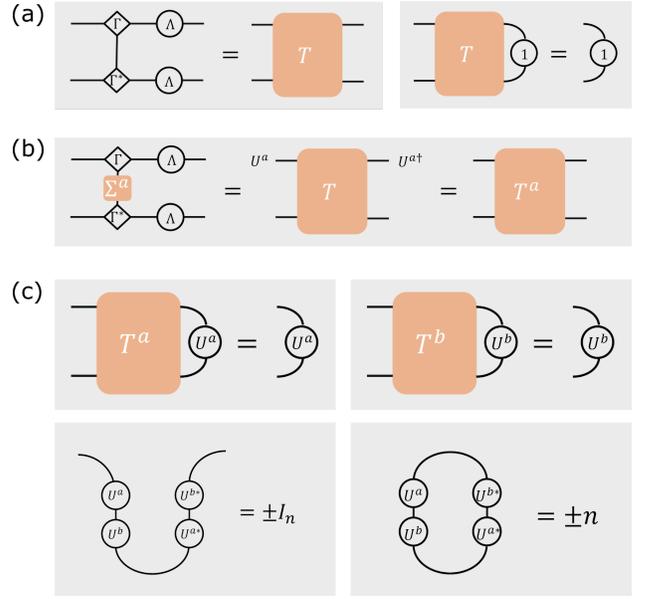}
    \caption{Brief illustration of the detection of SPT order. (a) iMPS and its transfer matrix $T$.  The corresponding leading eigenvector is the identity matrix, denoted by ``1.''. (b) iMPS transformed by a symmetry $\Sigma^a$ and the corresponding generalized transfer matrix $T^a$, which is given in terms of the original transfer matrix by a unitary transformation $U^a$ as shown.  (c) The leading eigenvector of $T^a$ is $U^a$. For two different elements $a$ and $b$ in the symmetry group $G = \mathbb{Z}_{2} \times \mathbb{Z}_{2}$, we can calculate $U^a U^b U^{a\dagger} U^{b\dagger}$ to determine whether the iMPS belongs to an SPT phase.}
    \label{fig:SPTdetection}
\end{figure}

For a given iMPS $\ket{\psi}$ and corresponding symmetry group, e.g. a $G=\mathbb{Z}_{2} \times \mathbb{Z}_{2}$ group and $g^a, g^b \in G$  the $\mathbb{Z}_2$ generators, we first apply the corresponding operators $\Sigma^a, \Sigma^b$ onto $\ket{\psi}$ to get the generalized transfer matrices $T^a, T^b$; then with $T^a$ and $T^b$ we calculate their largest eigenvalues $\eta^a, \eta^b$ and corresponding eigenvectors $U^a, U^b$.  If $|\eta^a|=|\eta^b|=1$ the iMPS indeed respects the symmetry $G$.  Furthermore, if $U^a U^b = - U^b U^a$, then $\ket{\psi}$ realizes a projective representation of the symmetry group; in contrast, $\ket{\psi}$ realizes a linear representation if $U^a$ commutes with $U^b$.  The method is summarized in Fig.~\ref{fig:SPTdetection}.

\begin{figure}
    \centering
    \includegraphics[width=0.85\linewidth]{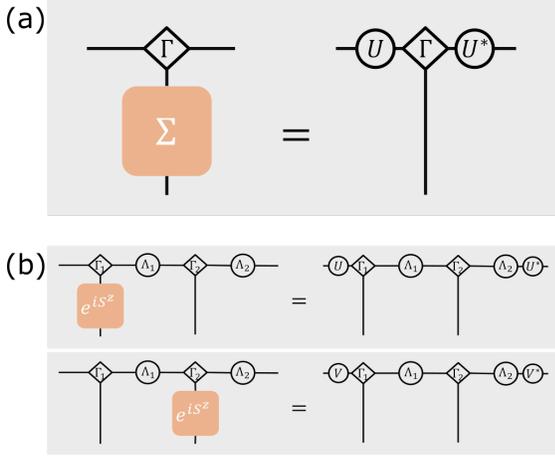}
    \caption{Modification of symmetry action on iMPS for model with two-site symmetries. (a) A usual translation-invariant iMPS under the transformation of its on-site group $G=\{ \Sigma \}$. (b) Action of symmetries on the ground state of the cluster model $H=\sum \sigma^x\sigma^z\sigma^x$, invariant under the translation by $2$ sites.}
    \label{fig:SPTdetection_cluster}
\end{figure}

Note that the above description takes for granted that the local symmetry applies on the system site by site.
However, we are often interested in systems where such translational symmetry is only partly respected.
For example, in the cluster model $H = \sum_i \sigma^x_{i} \sigma^z_{i+1} \sigma^x_{i+2}$, the symmetry $G=\mathbb{Z}_{2} \times \mathbb{Z}_{2}$ is implemented by two generators $Z_{odd} = \prod_{k} \sigma^z_{2k+1}$ and  $Z_{even} = \prod_{k} \sigma^z_{2k}$. We should therefore view each pair of neighboring sites together as a single unit cell, then apply the method discussed before.
Either of the two distinct groupings, $2k+1$ and $2k+2$ as a single unit cell or $2k$ and $2k+1$ as a single unit cell, is equally valid.  The effect of a symmetry transformation on an iMPS in this more general case is illustrated in Fig.~\ref{fig:SPTdetection_cluster}.

The case of the Kitaev ladder has further complications.  As established in Appendix~\ref{appendix:JW} above, we can order the sites so that the Hamiltonian appears to be translation-invariant with a two site unit cell; let the corresponding two-site translation symmetry be $\mathbb{T}_2$.  However, the global symmetries $\Sigma^Z_u$ and $\Sigma^Z_l$ then act differently on even and odd unit cells.  Explicitly, $\Sigma^Z_u \mathbb{T}_2 = \mathbb{T}_2 \Sigma^Z_l$, or intuitively $\mathbb{T}_2$ swaps $\Sigma^Z_u$ and $\Sigma^Z_l$.
We conclude that, although the Hamiltonian appears to have a two-site translation symmetry, in fact a four-site unit cell is needed when considering the action of symmetries.

\begin{figure}
    \centering
    \includegraphics[width=0.85\linewidth]{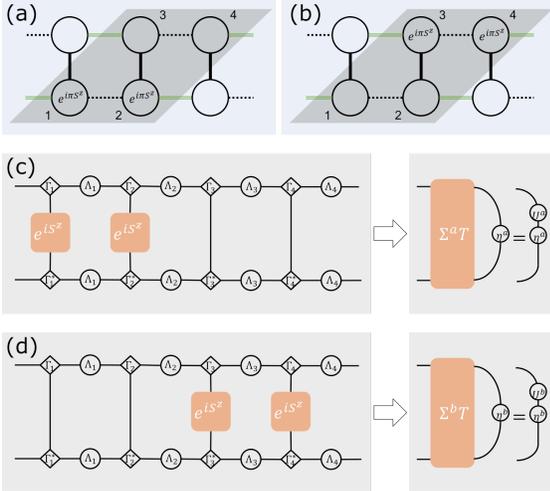}
    \caption{(a)(b)An intuitive choice of unit cell inspired by perturbation theory. (c)(d) With this choice of unit cell, $\Sigma^Z_l$ and $\Sigma^Z_u$ act, respectively, on the first two sites and the last two sites.}
    \label{fig:SPTdetection_KL}
\end{figure}

We then want to use the original iMPS, with a four-site unit cell, to detect the SPT order of the effective cluster model from perturbation theory.  In the $X$-limit or $Y$-limit where the cluster model arises, it seems natural to pick a unit cell that does not cut the dimers that form effective sites.  Such a choice is illustrated in the $Y$-limit in Figs.~\ref{fig:SPTdetection_KL}(a) and \ref{fig:SPTdetection_KL}(b).

\begin{figure}
    \centering
    \includegraphics[width=0.85\linewidth]{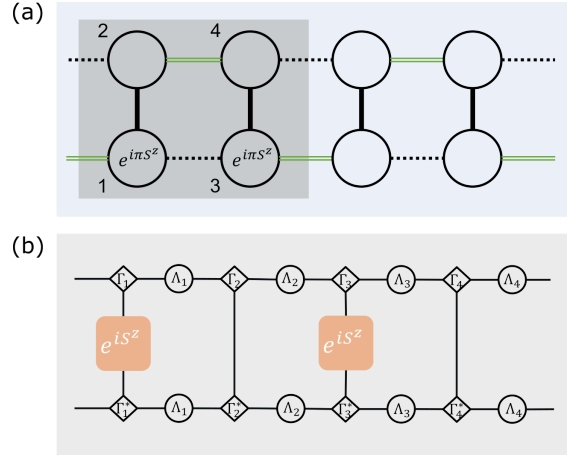}
    \caption{(a) An improperly selected  translationally invariant block for the same iMPS and (b) the corresponding local operators for the symmetry $\Sigma^Z_l$}
    \label{fig:SPTdetection_KL_wrong}
\end{figure}

What goes wrong if we shift the unit cell by one site, as in Fig.~\ref{fig:SPTdetection_KL_wrong}, so that the unit cell boundary cuts across a dimer?  This does not affect our measurement of the projective representation of $\Sigma_u^z \times \Sigma_l^z$.  However, if we perform the same measurement for the global symmetries $\Sigma^X$ and $\Sigma^Y$, we also naively find a signal of a projective representation, even though we know from perturbation theory that both symmetries map to the identity in the effective model and thus do not act nontrivially in the SPT phase.

\section{SOPs and SPT distinction}\label{appendix:SOPs}
Here we explain the construction of the SOPs of Fig.~\ref{fig:string}, used to distinguish the SPT-$x$ and SPT-$y$ phases of the Kitaev ladder.

In earlier explorations of string orders for distinguishing SPT phases and trivial phases~\cite{pollmann2012detection, morral2023detecting}, researchers defined decorated SOPs for detecting SPT phases.
The string bulk is given by (a local portion of) one of the symmetry operators, while two decorating operators that anticommute with the symmetry are added at the endpoints.  For Kitaev ladders, the construction of SOPs is more intricate because the local $D_n$ symmetries (see Sec.~\ref{subsection: symmetries}) constrain the allowed SOPs that can be constructed. 

To be more precise, following the same procedure of constructing SOPs for the cluster model, where one puts operators on even/odd positions, for the Kitaev ladder we can put $e^{i\pi S^z}$ (see Table~\ref{table:ladder_symmetry}) on either the upper leg or the lower leg~\cite{pollmann2012detection, morral2023detecting}.
However, using only $e^{i\pi S^z}$, at the two ends of each string there emerges anticommutation with $D_n$ which enforces the pure $Z$-strings (on a single leg) to have zero expectation value.
Therefore, it is necessary to decorate the end points of such $Z$-strings with either $e^{i\pi S^x}$ or $e^{i\pi S^y}$.
The SOPs from Fig.~\ref{fig:string} are:
\begin{equation}
     O_x = e^{i\pi S^x_1} \left( \prod_{n=1}^{N} e^{i\pi S^z_{4n-2}}e^{i\pi S^z_{4n-1}} \right) e^{i\pi S^x_{4N }}
\end{equation}
\begin{equation}
      O_y = e^{i\pi S^y_2} \left( \prod_{n=1}^{N} e^{i\pi S^z_{4n-3}}e^{i\pi S^z_{4n}} \right) e^{i\pi S^y_{4N-1}} 
\end{equation}

\begin{figure}
    \centering
    \includegraphics[width=0.95\linewidth]{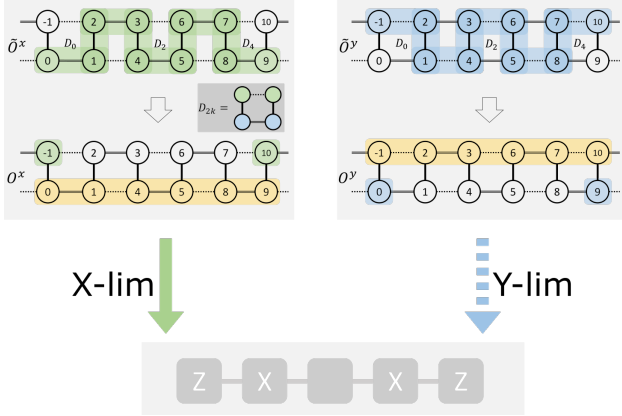}
    \caption{Illustration of the string order parameters and their equivalent definition after multiplication with the local symmetries. Two SOPs in different limits map to effective SOPs of cluster model in different ways, which highlights the phase distinction. Note that the capitalized $X$ and $Z$ in the lower part indicate the operators in the effective models.}
    \label{fig:refinedSOPs}
\end{figure}

${O}^x$ has a bulk built from $\Sigma^Z_u := \prod_{n=1}^{N} e^{i\pi S^z_{4n-2}}e^{i\pi S^z_{4n-1}}$ that commutes with all the symmetries, while the endpoint operators anticommute with $\Sigma^Z_l:=\prod_{n=1}^{N} e^{i\pi S^z_{4n-3}}e^{i\pi S^z_{4n}}$ and $\Sigma^X_u \Sigma^Y_l$. 
${O}^x$ can also be mapped by a two-site translation along the snake order to get an equivalent SOP built from $\Sigma^Z_l$ and with endpoints anticommuting with $\Sigma^Z_u$ and $\Sigma^X_l \Sigma^Y_u$; this shifted SOP is shown in Fig.~\ref{fig:refinedSOPs}).
Likewise, ${O}^y$ has endpoints that anticommute with $\Sigma^Z_u$ and $\Sigma^X_u \Sigma^Y_l$ (and has an equivalent two-site shifted version interchanging the roles of the upper and lower legs of the ladder).  

\begin{table}[htbp]
\renewcommand{\arraystretch}{1.2}
\begin{tabular}{ c  | >{\centering\arraybackslash}p{1.5cm}| >{\centering\arraybackslash}p{1.5cm} | >{\centering\arraybackslash}p{1.5cm} |>{\centering\arraybackslash}p{1.5cm} }
        \multirow{2}{*}{ }& \multicolumn{2}{c|}{$X_\textbf{even}$}  & \multicolumn{2}{c}{$X_\textbf{odd}$} \\  \cline{2-5}
        & rep1  & rep2  & rep1  & rep2  \\  \hline
   $X$-limit & $\Sigma^Z_u$ & $\Sigma^X_l\Sigma^Y_u$ & $\Sigma^Z_l$ & $\Sigma^X_u\Sigma^Y_l$ \\  \hline
   $Y$-limit & $\Sigma^Z_l$ & $\Sigma^X_l\Sigma^Y_u$ & $\Sigma^Z_u$ & $\Sigma^X_u\Sigma^Y_l$ \\  \hline
  \hline
\end{tabular}
 \caption{
 Pre-images in the original model of the $X_\text{even}$ and $X_\text{odd}$ symmetries of the effective cluster model from perturbation theory in the $X$- and $Y$-limits for half-integer spin.  ``rep1'' and ``rep2'' columns give two different elements of the original $\mathbb{Z}_2 \cross \mathbb{Z}_2 \cross \mathbb{Z}_2$ symmetry group, each of which map to $X_\text{even}$ or $X_\text{odd}$ in the specified anisotropic limit.} 
 \label{table:ladder_symmetry}
\end{table}

Note that, by multiplying local symmetries $D_2, D_4, \cdots$ with such SOPs, we can get the same $X$-strings or $Y$-strings as in \cite{PhysRevB.99.195112}.
For an illustration, see Fig.~\ref{fig:refinedSOPs}.

\section{SSB in isotropic limits\label{appendix:SSB}}
For all spin-$S$ values we consider, in the isotropic limit we find an SSB phase.  
As we show here, our results strongly suggest the existence of SSB at the isotropic point even in the large-$S$ limit.
Now we will 

One possible challenge in taking the large-$S$ limit is that, to find a smooth limit of the SSB order as $S$ increases, we would like to use the same order parameter in both cases, but the natural order parameter is different for half-integer and integer spins.  
Specifically, for half-integer spins it is most natural to detect the SSB phase using the operator from the $Z$-limit perturbation theory, $\expval{\ox \otimes \oy}$ (with the tensor product taken over the two sites in a rung of the ladder), but for integer spins that operator commutes with all symmetries and we instead use $O^{xy} = \expval{S^x \otimes S^y}$.  

Fortunately, the latter expectation value, $O^{xy}$, actually works in both cases.  The SSB phase breaks all the global symmetries except for $\Sigma^Z$, and indeed $O^{xy}$ commutes with $\Sigma^Z$ and anticommutes with the remaining global symmetries, and thus it is a valid order parameter for half-integer as well as integer spin.

\begin{figure}
    \centering
    \includegraphics[width=0.85\linewidth]{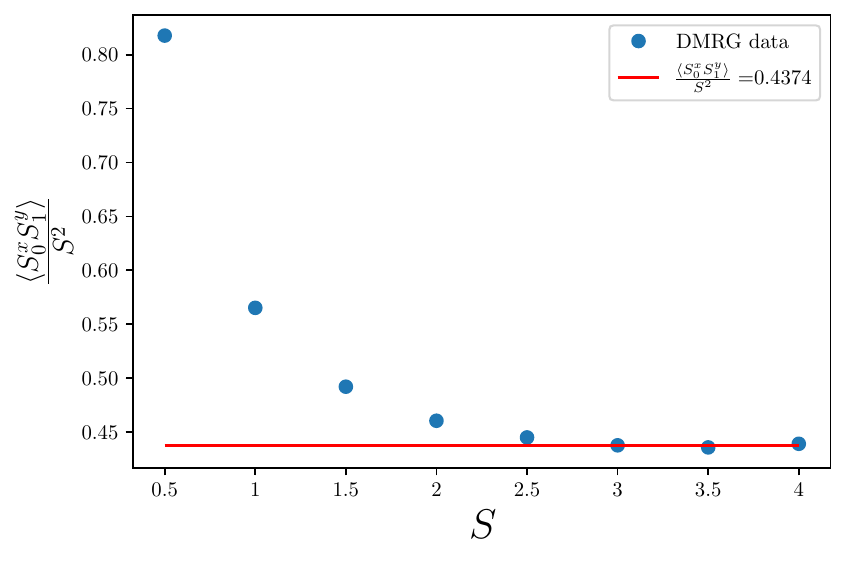}
    \caption{SSB order parameter $\langle S^x\otimes S^y\rangle/S^2$ of different spin-$S$ Kitaev ladders at the isotropic point ($K_x = K_y = K_z$).  Evidently the order parameter converges with increasing $S$, indicating that the SSB phase survives in the large-$S$ limit.}
    \label{fig:SSB_extrapolation}
\end{figure}

In Fig.~\ref{fig:SSB_extrapolation}, we plot the measured values for $O^{xy}$ for different spins ranging from $S=1/2$ to $S=4$.  We divide by the maximum possible value of $O^{xy}$, which is $S^2$, so that for any given $S$ a maximally symmetry-broken state would have value 1, and so that $O^{xy}$ will converge to a finite value in the limit $S\rightarrow\infty$.  

Evidently, the degree of symmetry breaking, as measured by $O^{xy}/S^2$, converges as $S$ increases, indicating that indeed the SSB phase will survive to all larger spin-$S$, beyond those we explicitly studied using DMRG.